\documentclass{aa}

\usepackage{graphicx}
\usepackage{placeins}
\usepackage{txfonts}
\usepackage{ulem}
\usepackage{pdflscape}
\usepackage{amsmath}
\usepackage[caption=false]{subfig}
\usepackage{float}
\restylefloat{table}
\usepackage{adjustbox}

\usepackage{siunitx}
\usepackage{multirow}
\usepackage{graphicx} 
\usepackage{xcolor}
\usepackage{mathabx}
\usepackage{hyperref}
\usepackage{tablefootnote}

\hypersetup{colorlinks,allcolors=blue}

\newcommand{\MEarth}{M$_{\mathrm{\oplus}}$\,}
\newcommand{\MJup}{M$_{\mathrm{Jup}}$\,}

\newcommand{\RSun}{R$_{\odot}$\,}
\newcommand{\Msun}{M$_{\odot}$\,}

\newcommand{\teff}{$T_{\rm eff}$\,}

\newcommand{\mdot}{$\dot{M}$\,}

\def\tablebib#1{\par\vspace*{2ex}
 \parbox{\hsize}{\leftskip0pt\rightskip0pt
 {\noindent\small\textbf{References.}~#1\par}}}

\begin{document}

   \title{The evolution of discs and the epoch of formation of giant planets around low-mass stars \thanks{raffaele.gratton@inaf.it}}

  \author{Raffaele Gratton\inst{1}, Elisabetta Rigliaco\inst{1} }

   \institute{INAF – Osservatorio Astronomico di Padova, Vicolo dell’Osservatorio 5, 35122 Padova, Italy\\ }

   \date{Received 29 Sept 2026 / Accepted 19 November 2026 }

\abstract{
Large-grain depletion in the inner discs of young stellar systems provides a potential observational signature of dust filtration by forming giant-planet cores. In a previous study, we found that $0.50\pm0.10$ of prominent protoplanetary discs around solar-type stars ($0.7$--$1.5$~\Msun) with ages of 2--8~Myr show evidence for a depletion of large grains inside the snow line. Such a depletion may arise when a sufficiently massive planetary core acts as a barrier to the inward drift of large grains, filtering them from the inner disc. If this interpretation is correct, the frequency of dust-filtration signatures should decrease towards lower stellar masses, in accordance with the lower occurrence rate of giant planets around low-mass stars. Extending our analysis to lower-mass stars, we found that only $0.09\pm0.09$ of discs around stars with masses of $0.30$--$0.58$~\Msun, corresponding to the progenitors of main-sequence early M-type stars, show evidence of grain filtration. This fraction is lower than that measured for solar-type stars and is consistent with the lower frequency of giant planets around M-type stars. We also found that discs exhibiting evidence of filtration tend to have relatively massive outer dust reservoirs, with dust masses exceeding $20$~\MEarth. Conversely, the absence of filtration signatures in lower-mass discs suggests that Neptune-mass and lower-mass planets, despite being common around low-mass stars, may generally be unable to filter large grains efficiently.
}


\titlerunning{The evolution of discs around low-mass stars stars}
\authorrunning{Gratton, R., and Rigliaco, E.}

   \maketitle

\section{Introduction}
\label{sect:Introduction}

Planet formation and disc evolution are closely related. The disc provides the raw material from which planets form and acts as the birthplace of planetary systems. The formation of planets may sculpt the disc, modifying its properties. It is generally thought that there are two main mechanisms for planet formation: disc instability \citep{Boss1997, Kratter2016} operating on a timescale of a few thousands years, and core accretion \citep{Safronov1972, Pollack1996, Mordasini2012, Armitage2020} operating  on a  slower timescale of millions of years. Disc instability is a top-down process and is expected to produce high mass companions at wide separation. Although the separation may change due to later migration to closer orbits, it is difficult to produce planets with mass ratios $q\leq 0.001$ with respect to their host stars through the disc instability process \citep{Kratter2016}. For this reason, core accretion is the favoured mechanism for the formation of giant planets in the Solar System and for exoplanets, too.

A peculiar feature of the core accretion mechanism is that once the cores of giant planets have formed (1--2~\,Myr after the initial formation of the stars), the inward drift of large grains \citep{Weidenschilling1977} from the outer to the inner (long living) disc is likely inhibited, creating two separate reservoirs of grains \citep{Paardekooper2006, Rice2006, Kalyaan2023, Drazkowska2019, Houge2026}. This process is called dust filtration. The segregation between non-carbonaceous and carbonaceous chondrites in the Solar System may be a manifestation of this phenomenon \citep{Kruijer2017, Kleine2020}. \citet{Weber2018} showed that only particles larger than a certain size can effectively be stopped from moving inward for a given diffusivity. Their calculations indicate that an appreciable effect occurs for $r_{\rm grain}\geq 0.03$~cm and is related to a planet to star mass ratio $q\approx 5\times 10^{-4}$, which is about half of the Jupiter mass for a solar type star. On the other hand, dust grains undergo continuous collisions, leading to fragmentation (break-up) and coagulation (sticking) of the interacting particles, depending on the material properties and collision velocities \citep{Birnstiel2024}. Following these phenomena is then important to establish the ability of a planet gap to affect the size distribution of grains in the inner disc \citep{Zhu2012}. Updated hydrodynamical simulations investigating dust filtration by planet-induced gaps are presented by \citet{Pfeil2025}. These authors performed two-dimensional multifluid hydrodynamic simulations of planet--disc systems that include dust coagulation and fragmentation. They showed that the efficiency of dust filtration is highly sensitive to grain evolution and, consequently, to disc properties such as turbulent diffusivity. Their results indicate that a planet with a mass of 0.6~\MJup likely produces a significant depletion of large grains in the inner disc. In contrast, for planets with masses of 0.1~\MJup -- comparable to the expected mass of Jupiter's core after (1--2~Myr) -- dust filtration is generally insufficient to overcome grain coagulation, which continuously replenishes the population of large grains in the inner disc.
Additionally, the impact of dust coagulation was also investigated by \citet{Drazkowska2019} who employed 2D hydrodynamic simulations of protoplanetary disks, showing that significant leakage of small grains can occur if fragmentation operates at the outer gap edge.

Pressure bumps leading to dust filtration might actually be present even in the absence of forming  planets (see \citealt{Chang2023}). For instance, \citet{Charnoz2021} found that when the vertically averaged viscosity index is a decreasing function of the surface density, then the release of water vapour at the snow line decreases the sound velocity, and a pressure bump may appear given a constant inflow of icy pebbles with a ratio of the pebble influx to gas influx $>0.6$~for a power-law disc with a 1\% ice-to-gas ratio, and $>1.8$~for a disc with an ice-to-gas ratio $\sim 0.3$\%. On the other hand, dust traps may favour the formation of giant planets (see, e.g., \citealt{Izidoro2022}). Then they may still be related to the phenomenon of dust filtration.

For several reasons, the region around and just beyond the snow line is considered a particularly favourable site for giant-planet formation \citep{Drazkowska2017}. Tracking the evolution of the dust size distribution both within and beyond the snow line is therefore crucial to quantify the efficiency with which planetary gaps filter dust grains. In a previous paper (\citealt{Gratton2026}, hereafter Paper I), we investigated the properties of dust grains in the inner and outer regions of protoplanetary discs, separated by a boundary approximately corresponding to the snow line. We performed a simultaneous modelling of the spectral energy distributions (SEDs) and ALMA dust-continuum images for a sample of 33 protoplanetary discs around solar-type stars, with stellar masses ranging from 0.7 to 1.5~\Msun. The sample was selected to include systems for which accurate stellar ages could be derived from dynamical masses and evolutionary tracks. We found that the grain size distributions in the inner (interior to the snow line) and outer regions of the disc are statistically indistinguishable for systems younger than 2~Myr. In contrast, among discs with ages between 2--8~Myr, approximately half exhibit a significant depletion of large grains in the inner disc. As mentioned above, this is the expected consequence of the existence of cores with several tens of Earth masses whose formation requires 1--2~Myr \citep{Paardekooper2006, Rice2006, Zhang2018, Weber2018, Pfeil2025}. In contrast, the lack of evidence of dust filtration in very young discs would need a specific explanation for alternative mechanisms. Although the sample we considered is biassed because we selected stars with prominent long-living discs, the fraction of long-living discs showing evidence for grain segregation agrees with the frequency of Jupiter-like planets observed in young associations, which are the likely progeny of the sparse star forming regions in the Solar neighbourhood. 

However, it is also well known that the frequency of giant planets is lower around low-mass stars with respect to solar type stars \citep[e.g.][]{Johnson2010, Bryant2023}, likely because the mass of the protoplanetary disc (superlinearly)  depends on the mass of the star \citep{Andrews2013, Mulders2018}, and cores forming in discs around low-mass stars rarely achieve the critical mass for gas accretion that would transform them into giant planets \citep{Greaves2010}. On the other hand, low-mass stars host lower mass planets (Neptune-like or even planets of lower mass) even more frequently than solar-type stars do \citep{Mulders2015, Mulders2018}. If dust segregation is indeed related to the formation of massive cores, but not chains of smaller mass planets, we might then expect a lower frequency of discs showing evidence of dust segregation when observing stars of lower mass. To test this prediction, we employ the same methodology adopted in Paper I to investigate here a sample of 18 protoplanetary discs around low-mass stars (mass $<0.7$~\Msun). Section \ref{sect:data} describes the data that we used. Section \ref{sect:methods} outlines the methods used to analyse these data. Section \ref{sect:results} presents the results that we discuss in Sect. \ref{sect:discussion}. We draw conclusions in Sect. \ref{sect:conclusions}.

\section{Data}
\label{sect:data}

The stellar sample analysed in this paper is listed in Tables~\ref{tab:uv_data} and \ref{tab:input_data}. It comprises 18 stars, all but one of which have dynamical masses below $\sim 0.7$ \Msun. Most of the stars in this sample (14 out of 18) have masses lower than $0.58$~\Msun, with the lowest stellar mass being $0.3$ \Msun. Based on the mass values associated with the spectral types reported by \cite{Pecaut2013}\footnote{\url{https://www.pas.rochester.edu/~emamajek/EEM_dwarf_UBVIJHK_colors_Teff.txt}}, these stars are expected to be progenitors of early main-sequence M-type stars. The sample therefore extends the range of stellar masses explored in Paper I.

The sample comprises stars with dynamical masses derived from the Keplerian rotation of their circumstellar discs, as traced by molecular-line observations obtained primarily with ALMA. As in Paper I, we exploited the availability of dynamical mass estimates to derive relatively accurate ages for individual stars using the age–radius relation provided by the evolutionary models of \citet{Baraffe2015}. The dynamical masses were compiled from several sources: \cite{Simon2019, Tsukagoshi2019, Braun2021, Trapman2025, Longarini2025}. In addition, given the limitations of the model adopted in our analysis, as described in Sect.~\ref{sect:methods} and in Paper I, we restricted the sample to discs with inclinations $i<70^\circ$ and without bright companions within 2 arcsec, which is roughly 300 au considering the distance of the programme stars.

As in Paper I, data used in this paper are the spectral energy distribution (SED) and high-resolution images of dust continuum emission from ALMA bands 6 and 7. The SED were obtained using the values provided by the SIMBAD VizieR Photometry viewer \citep{Ochsenbein2000}\footnote{\url{https://cds.unistra.fr/news/2014/11/13-photometry-viewer/}}. We generally gave preference to values provided by {\it Gaia} DR3 \citep{GaiaCollaboration2022}, 2MASS \citep{Skrutskie2006}, {\it WISE} (WISE All-Sky Data Release, \citealt{Wright2010, Cutri2012}), {\it Spitzer} \citep{Spitzer2021}, {\it Herschel} (PACS: \citealt{Ribas2017};  Herschel/SPIRE point source catalogue (HSPSC): \citealt{Herschel2024}), Submillimetre Common-User Bolometer Array (SCUBA) at the James Clerk Maxwell Telescope  \citep{Mohanty2013}, and ALMA, but also considered additional near UV and $B$-band photometry from the SkyMapper Southern Sky Survey DR1.1 \citep{Wolf2018}. 

We also gathered space UV data that are crucial for obtaining sensible values for the accretion rates (see Table~\ref{tab:uv_data}). These were obtained from an inspection of data obtained using the Cosmic Origins Spectrograph (COS) and the Space Telescope Imaging Spectrograph (STIS) on board of the Hubble Space Telescope as available in the Barbara A. Mikulski Archive \footnote{\url{https://mast.stsci.edu/portal/Mashup/Clients/Mast/Portal.html}}. In addition, we considered data from International Ultraviolet Explorer (IUE) \citep{Beitia-Antero2016}, Galaxy Evolution Explorer (GALEX) \citep{Bianchi2011}, XMM-Newton \citep{Page2023}, and UVOT \citep{Yershov2014}. Overall, space UV data were available for seven stars (CY Tau, Sz 114, CX Tau, HM Lup, DM Tau, Sz 111, and CW Tau). Tables listing the final data used in the definition of the SED will be made available in the VIZIER database at the Centre de Données astronomiques de Strasbourg (CDS).

The collected photometry was corrected for interstellar absorption using the law by \citet{Cardelli1989} assuming a ratio of $R_{V}=3.1$~mag between total and selective absorption. The value of the total interstellar absorption in the $V$--band for each star was obtained from the difference between the observed $B_P-R_P$ colour in {\it Gaia} DR3 and that expected for stars of the same spectral type according to the tables by \citet{Pecaut2013}\footnote{\url{https://www.pas.rochester.edu/~emamajek/EEM_dwarf_UBVIJHK_colors_Teff.txt}}, assuming $E(B_P-R_P)=0.457~A_V$ as given by the law of \citet{Cardelli1989}.

The ALMA Science archive\footnote{\url{https://almascience.nrao.edu/aq/}} was used to collect high-resolution ($\rho<0.15$~arcsec) data in bands 6 and 7 reduced with {\tt CASA} software\footnote{\url{https://casa.nrao.edu/index.shtml}}. The datasets containing the continuum image used in our analysis are listed in Table~\ref{tab:alma_data}. Archived images are in physical flux units and have a variable scale and resolution, as defined by the beam parameters contained in the header of the fits files. These were considered in our analysis.

\section{Methods}
\label{sect:methods}

\subsection{Model description}

We used the data described in Sect.~\ref{sect:data} to derive the main stellar and disc properties using the same simplified model described in detail in Paper I, where it was applied to a sample of solar-type stars. Here, we provide a brief summary of the model.

Our simplified parametric model simultaneously fits the SEDs and high-resolution ALMA continuum images. Compared with the radiative-transfer models commonly employed to reproduce SEDs and ALMA observations (e.g. \citealt{Woitke2019}), our approach adopts a substantially simplified description of both the three-dimensional disc structure and the dust-grain properties. This reduces the computational cost and number of free parameters, making the model well suited to the analysis of large samples. In particular, the range of parameters typically used to describe the chemical composition and internal structure of dust grains is reduced to a single parameter, the grain size. This simplified treatment enables an efficient investigation of disc evolution across large samples, but the resulting parameters should be interpreted primarily in a qualitative rather than quantitative sense.

Our code requires as input information the dynamical mass $M_\star$ for the star, the SED, a high resolution (Full Width at Half Maximum FWHM$<0.15$~arcsec) ALMA continuum image in bands 6 or 7 for each star, the parallax $\pi$ of the star (from Gaia DR3, \citealt{Gaia2023}), the interstellar absorption $A_V$ in the $V$ band in its direction (obtained as described in the previous section), and the inclination $i$ and position angle $PA$ of the disc (collected from the literature, see Table \ref{tab:input_data}). The output parameters provided by the fit to the observational data are the stellar effective temperature (\teff) and radius ($R_*$), the accretion rate (\mdot), the masses of the inner and outer portions of the disc ($M_{\rm inner}$) and ($M_{\rm outer}$), the inner and outer radii of the inner disc ($a_{\rm min, inner}$) and ($a_{\rm max, inner}$), and the maximum radius of the grains in the inner disc ($R_{\rm max,grain,inner}$). 

Given a set of input data and parameters, the code computes both the SED and the images of the disc in the continuum at the appropriate wavelength, taking into account the inclination and position angle of the disc, convolved for the beam as read from the header of the FITS files containing the ALMA images.

The model includes three contributors to the observed SED:\\
{\it (i)} Stellar photosphere -- modelled using Kurucz stellar atmosphere models\footnote{\url{https://www.stsci.edu/hst/instrumentation/reference-data-for-calibration-and-tools/astronomical-catalogs/kurucz-1993-models}} for \teff$>3500$~K and Phoenix models \citep{Husser2013} for lower \teff. We adopt the Kurucz model atmospheres for stars with \teff$>3500$~K, consistent with Paper I. For lower effective temperatures, we instead use the PHOENIX models, as stars with $M_\star<0.7$~\Msun often have \teff$<3500$~K, below the lower limit of the Kurucz grid. We assumed surface gravities of $\log{g}=4.0$\ in both cases. The stellar effective temperature and radius are outputs of the model.\\
{\it (ii)} Accretion -- represented as a 10,000 K blackbody region produced by accretion shocks on the stellar surface. The accretion luminosity is derived from the mass accretion rate \mdot that is an output of the model considering the stellar radius and mass. We assumed that half of the contraction energy is radiated (as expected for viscous discs) and that the ratio between the inner edge (of the gas) disc and the star radius is four, as in \citet{Alcala2017}.\\
{\it (iii)} Dusty disc emission -- The disc emission is assumed to arise solely from thermal dust emission, neglecting gas emission and scattered stellar light. The disc is represented as a two-dimensional, radially symmetric structure divided into inner and outer regions. The boundary between the two regions is defined by the outer radius of the inner disc, initially set to the snowline and subsequently adjusted to reproduce the observations; this radius is therefore a free parameter and an output of the model.
The surface density of the inner disc, which is unresolved in the ALMA images, is assumed to follow an $1/a$ radial profile, where $a$ is the distance from the star. In the outer disc, the surface density profile is constrained by the flux distribution along the disc major axis in the ALMA observations. The local disc temperature is determined assuming radiative equilibrium with the stellar and accretion luminosities. We neglect the vertical disc structure and shadowing effects, which may result in overestimated temperatures in thick discs.
The dust grain-size distribution is described by a fixed power law with a minimum grain size of $0.1~\mu$m and independent maximum grain sizes in the inner and outer discs. The dust is composed of silicate and ice grains with fixed abundances, densities, and albedos. Grain survival is determined by sublimation temperatures of 1500~K for silicates and 170~K for ices. Dust opacities are calculated using the simplified prescription of \citet{Mordasini2014}, in which the opacity depends only on the ratio of grain size to wavelength. This treatment neglects the detailed dependence of the opacity on grain composition and internal structure, but avoids introducing additional free parameters and is therefore appropriate for the simplified parametric approach adopted here.

\subsection{Model's caveats}

The assumptions used in the proposed model are intentionally simplified. Quantities such as disc masses should therefore be interpreted cautiously. However, because the same assumptions are applied consistently across all stars, comparative trends, such as how disc properties evolve with age, are still considered meaningful.

An important caveat in our model is that, given the observables considered, we determined the mass in dust rather than the total mass of the disc, which is mainly due to gas. We prefer to use total masses (of the inner and outer discs) as free parameters because this allows a simpler description of the surface density with distance from the star, as explained in Paper I. Note that using this approach, the gas to dust ratio is always 100 in the outer disc, while it may be much higher in the inner disc, depending on the distance from the star (because only silicate grains are considered within the snow line) and on the actual value of $R_{\rm max,grain,inner}$.

Finally, we should warn the reader that our procedure works well insofar each component considered (stellar photosphere, accretion region, inner and outer disc) dominates the SED in some wavelength range. Some parameters are conversely poorly determined if this does not occur. As an example, if the contribution due to accretion dominates over that from the stellar photosphere even at the optical wavelengths, which happens when the veiling is very high, the parameters defining the photosphere (stellar radius and effective temperature) are poorly constrained. Among the stars considered in this study, this is the case for CW Tau (see Appendix \ref{app:input_data}). For this star, we adopted the age (and then the corresponding radius)  from the association to which the star belongs.

An additional simplification of our model, which, however, does not affect the results, is that the transition radius between inner and outer disc is sometimes smaller than the ALMA angular resolution beam, implying that the central intensity includes a contribution from the outer disc.  
To assess whether the finite angular resolution of the ALMA observations could introduce a systematic bias in the inferred values of \(R_{\rm max,grain,in}\), we examined their dependence on the ratio between the ice-line radius, \(a_{\rm ice}\), and the spatial resolution of the ALMA images. This ratio is generally smaller for lower-mass stars, as expected. However, we find no significant correlation between \(R_{\rm max,grain,in}\) and \(a_{\rm ice}/\mathrm{resolution}\), indicating that the inferred grain sizes are not systematically affected by the angular resolution of the observations.

\subsection{Treatment of noise} 

\begin{figure}[htb!]
    \centering
    \includegraphics[width=\columnwidth]{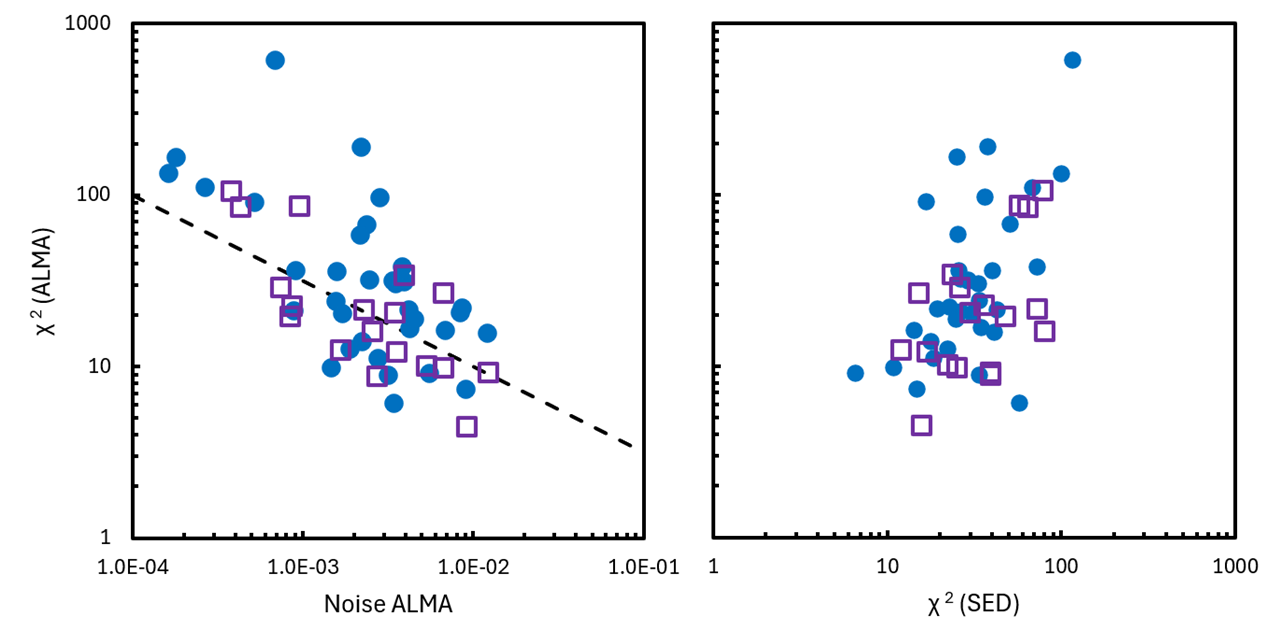}
    \caption{Left: $\chi^2$ obtained for the best fit of the ALMA images as a function of the noise in these images. Dashed line corresponds to the assumption that actual noise depends on the square root of the signal, as expected for photon noise. Right: $\chi^2$ obtained for the best fit of the ALMA images as a function of $\chi^2$ for the SED. Purple open squares are results for the discs analysed in this paper, blue filled circles are for those analysed in Paper I.}
    \label{fig:chisquare}
\end{figure}

Keeping in mind the qualitative nature of our analysis and the simplifications adopted to reduce the number of free-parameters, we comment here on how we treated the noise in our study. The noise in the photometric points considered in our analysis of the SED is the maximum between the error provided by the original source and an arbitrary value of 5\%. Since the sources may be variable, this avoids attributing too much weight to some of the photometric points.

The noise in the ALMA images was estimated as the standard deviation measured in regions far from the central star. Figure~\ref{fig:chisquare} compares the $\chi^2$ values obtained for the best fitting ALMA images with those expected from the estimated image noise (left panel) and with the $\chi^2$ values of the best fitting SED models (right panel). The left panel shows that lower image noise is associated with higher $\chi^2$ values for the best-fitting models. The trend suggests that our procedure underestimates the variance in the ALMA images, likely because the noise is not independent of the signal. The dashed line shows the expected relation if the noise scales with the square root of the signal, as expected for photon noise. This relation provides a good description of most of the sample, with the main outliers being discs around stars more massive than those considered in this paper, which occasionally exhibit strong asymmetries (see Paper I for a discussion of these cases). Nevertheless, we consider the assumption of signal-independent noise to provide a more appropriate overall weighting of the ALMA images. If the noise were assumed to increase with the signal, the large number of low-intensity pixels would dominate the fit, while the relatively few high-intensity pixels would receive less weight. Adopting a constant noise level effectively assigns weights to individual pixels that are proportional to their signal, ensuring that the brightest regions, which contain much of the constraining information, have an appropriate influence on the fit.
The right panel of Fig.~\ref{fig:chisquare} shows a correlation between the $\chi^2$ values obtained for the best-fitting ALMA images and those obtained for the SED. This correlation is expected, as the fitting procedure optimises the sum of the two contributions to the total $\chi^2$.

Given the approximate treatment of the observational uncertainties adopted in this study, the uncertainties reported for the derived physical parameters should be regarded as indicative rather than as rigorous statistical confidence intervals. They provide an estimate of the robustness of the inferred parameters within the adopted modelling framework, rather than formal measurement errors.

\subsection{Optimisation of the model's solutions}

\begin{table}[t]
\caption{Output parameters and the ranges for random extraction.}
\centering
\begin{tabular}{lcc}
\hline
\hline
Parameter            & Range      & Unit \\
\hline
\teff                & $\pm$100   & K   \\
$R_*$          & $\pm$10\%  &  \RSun   \\
$\log{\dot{M}}$      & $\pm$0.4   & M$_\odot\,\rm{yr}^{-1}$ \\
$M_{\rm inner}$     & $\pm$20\%  &   \Msun  \\
$M_{\rm outer}$      & $\pm$20\%  &   \Msun  \\
$a_{\rm min, inner}$ & $\pm$10\%  &  au   \\
$a_{\rm max, inner}$ & $\pm$10\%  &  au   \\
$\log{R_{\rm max,grain,inner}}$ & $\pm$0.6& cm  \\
\hline
\end{tabular}
\label{tab:range}
\end{table}

Optimisation of free parameters was achieved by minimising the overall normalised $\chi^2$ value. The code allows for both manual optimisation (setting the number of random iterations to zero) and an arbitrary number of random iterations. They were obtained by considering random sets of values of the free parameters with uniform distributions within a given range from the original values that depend on the quantity considered (see Table \ref{tab:range}). Whenever the $\chi^2$ value obtained in this way was lower than that obtained with the original value, the best set of values is saved and used in the next iterations. The random walk allows progression towards a minimum solution. In practice, manual optimisation was used as a first step to provide the best centre of the images and a first approximation of the parameters. Subsequently, an automatic optimisation was launched until there was no further reduction of the value $\chi^2$ in the last 100 iterations. Finally, we again launched the automatic procedure by halving the ranges of variations used in the random extractions with respect to the values listed in Table \ref{tab:range}. Again, we stopped the procedure only after there was no reduction of $\chi^2$ in the last 100 iterations. The procedure was completed by estimating uncertainties on the best fit parameters derived from all solutions whose $\chi^2$ value was higher than the best value by less than 10\%. The parameters obtained through our analysis are listed in Tables \ref{tab:output_data} and \ref{tab:derived_data}.

\subsection{Covariances}

The covariance between the various quantities derived from our analysis can be examined using corner plots. 
Examples of these corner plots are given Appendix~\ref{sec:corner_plot}, in Fig.~\ref{fig:corner_plot}. To derive them, we used 10,000 additional random extractions of the parameters around the best values from uniform distributions over the ranges given in Table \ref{tab:range}. We then evaluated the probability for each of these models as $p=\exp{(-\chi^2/2)}$. We plotted isocontours containing 68.3\% (1$\sigma$) and 95.5\% (2$\sigma$) of the probabilities in the posterior distributions.

As discussed in Paper I, we may describe the distribution of grain sizes in the inner disc (different from what is assumed in the outer disc) by considering either the maximum value $R_{\rm max, grain}$ for the grain size or a different slope $\beta$ of the grain distribution. As expected, we found very similar best $\chi ^2$ values using the two approaches. There is a clear correlation between $\beta$ and $R_{\rm max, grain}$ for the best solutions. Optimising $\beta$ rather than $R_{\rm max, grain}$ results in different values of the inner disc mass. The ratio of the values for the inner disc mass obtained using $\beta$ and $R_{\rm max, grain}$ is strongly correlated with the value of $\beta$. The other quantities (in particular the stellar ages) are not affected by this choice, although the best values may be slightly different due to our use of a Monte Carlo approach.

\section{Results}
\label{sect:results}

\subsection{Selection bias of the sample} 
\label{sect:bias}

The stars included in our sample host relatively massive discs, which may introduce a selection bias in our results. To assess the magnitude of this bias, we consider the complete sample of stars in Taurus with spectral types earlier than M8.5, extending into the brown-dwarf mass regime, for which submillimeter continuum fluxes at 0.89~mm were obtained by \citet{Andrews2013}. Comparable data at 0.89 and 1.33~mm are available for a complete sample of stars in Lupus with spectral types earlier than M5.5, corresponding to stellar masses down to approximately 0.12~\Msun\ \citep{Ansdell2016, Ansdell2018}. Therefore, these surveys extend well below the lower mass limit considered in the present study. In addition, Taurus and Lupus account for a substantial fraction of the sources in our sample, which makes them suitable for assessing the selection bias affecting the overall sample.
The discs included in our sample typically have continuum fluxes $>25$~mJy at 0.89~mm and $>10$~mJy at 1.33~mm. Therefore, we estimated the fraction of discs in Taurus and Lupus exceeding these flux thresholds using the samples of \citet{Andrews2013}, \citet{Ansdell2016}, and \citet{Ansdell2018}, considering stars with masses in the ranges 0.30--0.58 and 0.7--1.4~\Msun. To ensure a consistent comparison, we considered the systematic differences between the dynamical masses adopted in this study and in Paper I and the stellar masses reported by \citet{Andrews2013} and \citet{Ansdell2016}. Specifically, when using the isochrones of \citet{Baraffe2015}, our dynamical masses are, on average, 0.67 times those reported by \citet{Andrews2013}, and 1.19 times those reported by \citet{Ansdell2016} and \citet{Ansdell2018}.
After applying these corrections, we found that 29 out of 66 discs around progenitors of main-sequence early M-type stars ($0.3<M_\star<0.58$~\Msun) in Taurus and Lupus have 0.89-mm fluxes above 25~mJy, corresponding to $44\pm6$\%. For comparison, 36 out of 50 discs around progenitors of solar-type stars ($0.7<M_\star<1.4$~\Msun) exceed this threshold, corresponding to $72\pm6$\%. These fractions indicate that our sample is subject to a selection bias towards relatively massive discs, and that this bias is stronger for the low-mass stars considered in this paper than for the solar-type stars considered in Paper I. We use this estimate of the selection bias when interpreting our results in Sect.~\ref{sect:discussion}.

\subsection{Relation between input and output parameters}

\begin{figure*}[!t]
    \centering
    \includegraphics[width=0.92\textwidth]{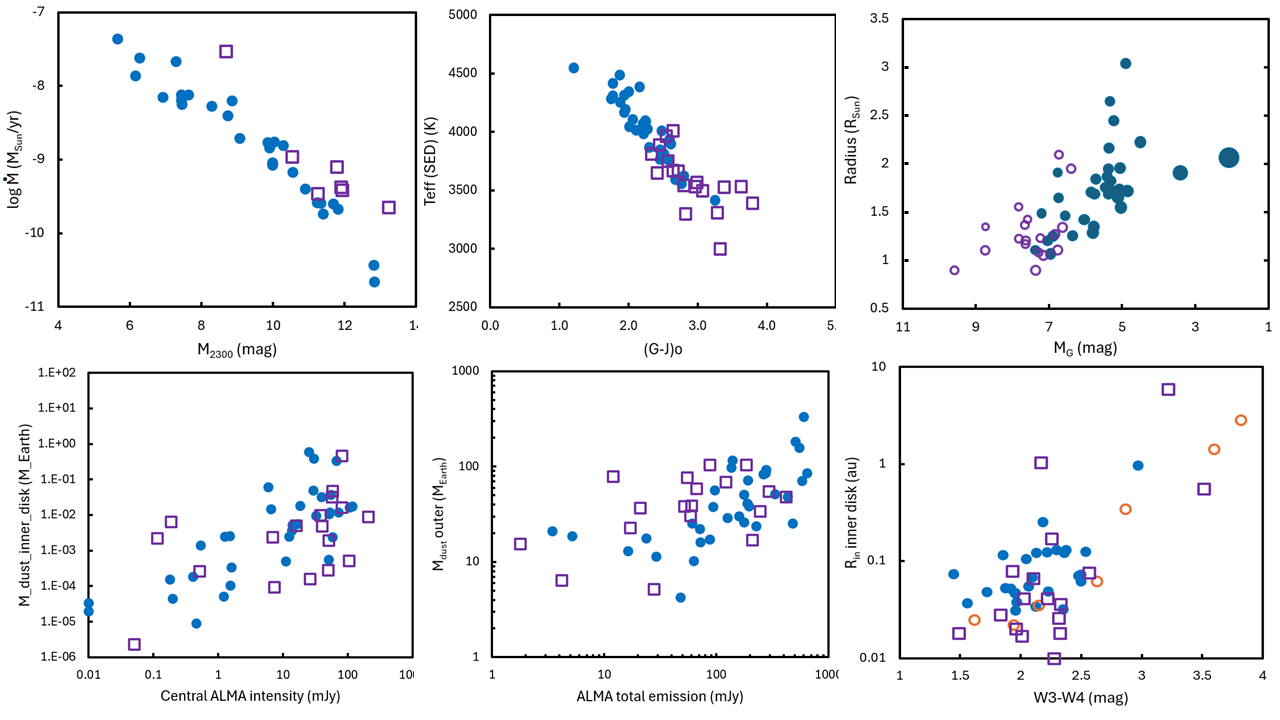}
    \caption{Comparisons between some input and output parameters used in our analysis. In all panels, purple open squares are results for the discs analysed in this paper, blue filled circles are for those analysed in Paper I. Upper left panel: absolute UV magnitude at 2300~\AA\ and accretion rate $\dot{M}$; upper central panel: dereddened $G-J$ colour and effective temperature; upper right panel: absolute $G$ magnitude and radius (in this panel the diameter of the symbols is proportional to the effective temperature of the star); lower left panel: intensity of the ALMA image at the star position and dust mass of the inner disc; lower central panel: total flux measured by ALMA and dust mass in the outer disc; lower right panel: inner radius of the inner disc and $W3-W4$ colour according to WISE photometry (in this panel, open symbols are binaries).}
    \label{fig:param_comparison}
\end{figure*}

Although the model optimisation is performed using the total $\chi^2$, individual model parameters are primarily constrained by specific observational quantities. Some of the resulting parameter correlations are shown in Fig.~\ref{fig:param_comparison}. 
A tight correlation is found between the mass accretion rates and the UV flux, which can be characterised by the absolute magnitude at 2300~\AA, derived from the observed flux after correction for interstellar extinction and distance (upper left panel of Fig.~\ref{fig:param_comparison}). For several stars, this magnitude is approximate, as it is estimated by interpolating between fluxes measured at other UV wavelengths. The offset between the low-mass stars analysed in this work and the solar-type stars from Paper I can be attributed to their lower surface escape velocities, which result in lower UV fluxes for a given mass accretion rate. Since the accretion luminosity scales with the mass accretion rate and the square of the escape velocity, the mass accretion rates inferred by the model are expected to scale approximately as the inverse square of the escape velocity. On average, the stars with available UV magnitudes in this sample have escape velocities that are 59\% of those of the stars in Paper I. We therefore expect an offset of a factor of $\sim3$ (0.46 dex) in the inferred mass accretion rates, consistent with the observed difference.

The stellar effective temperatures are constrained by the optical SED shape, here characterised by the $G-J$ colour; similar results are obtained using $B_P-R_P$. As expected, this yields a tight correlation (central upper panel of Fig.~\ref{fig:param_comparison}). Because the effective temperature range of pre-main-sequence stars within the mass range considered here is relatively narrow, the stellar radius is primarily determined by the absolute magnitude. This is illustrated in the upper right panel of Fig.~\ref{fig:param_comparison}, where the symbol size indicates the residual dependence on effective temperature. The offset between the low-mass stars (purple symbols) and the solar-type stars (blue symbols, Paper I) reflects the mass--temperature correlation expected for pre-main-sequence stars, as illustrated, for example, by the isochrones of \citet{Baraffe2015}.

\begin{figure}
    \centering
    \includegraphics[width=0.8\columnwidth]{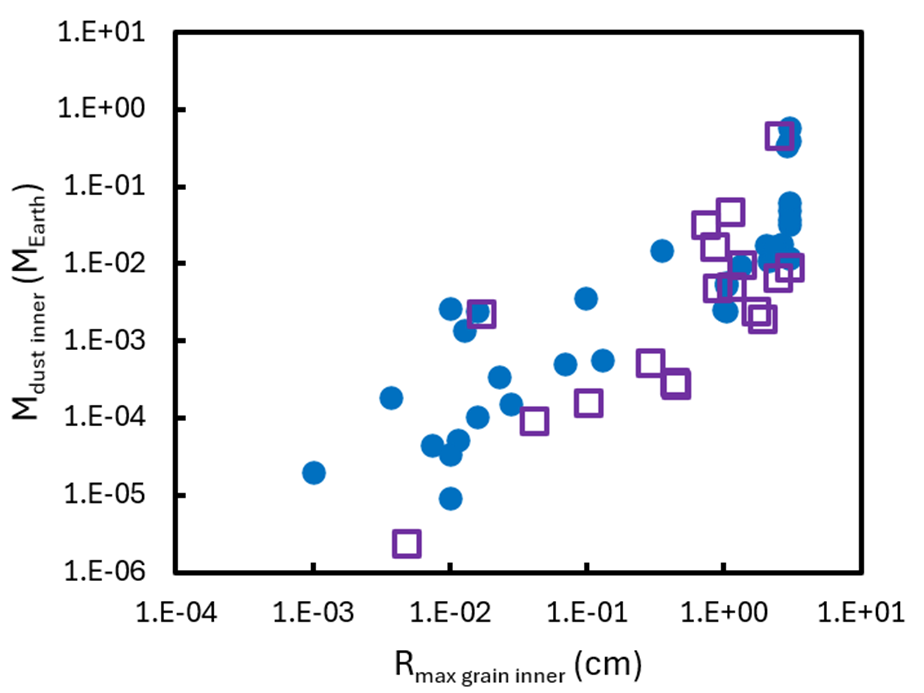}
    \caption{Correlation between the maximum radius for grains and dust mass in the inner disc. Purple open squares are results for the discs analysed in this paper, blue filled circles are for those analysed in Paper I.}
    \label{fig:rmax_grain}
\end{figure}

The masses of the inner and outer disc regions are closely linked to the ALMA continuum flux. The inner disc mass correlates with the central flux, but this region contributes only marginally to the total emission, which is instead more closely correlated with the dust mass in the outer disc, where most of the dust resides (lower-left and lower-central panels of Fig.~\ref{fig:param_comparison}). The inner disc mass is primarily determined by the maximum grain size required to reproduce the low central ALMA continuum emission while matching the mid-IR SED (Fig.~\ref{fig:rmax_grain}), indicating a depletion of large grains in the inner disc.

Finally, the inner disc radius sets the peak of the dust temperature and is therefore strongly correlated with the thermal-IR colours, particularly the $W3-W4$ WISE colour, which traces the 12-to-22~$\mu$m flux ratio\footnote{\url{https://www.astro.ucla.edu/~wright/WISE/passbands.html}} (lower-right panel of Fig.~\ref{fig:param_comparison}).

In conclusion, the main stellar and disc properties are strongly correlated with simple observables, enabling a straightforward physical interpretation and supporting the robustness of the derived parameters.

\subsection{Comparison with results from previous studies}

\subsubsection{Stellar effective temperatures}

\begin{figure}[htb!]
    \centering
    \includegraphics[width=0.8\columnwidth]{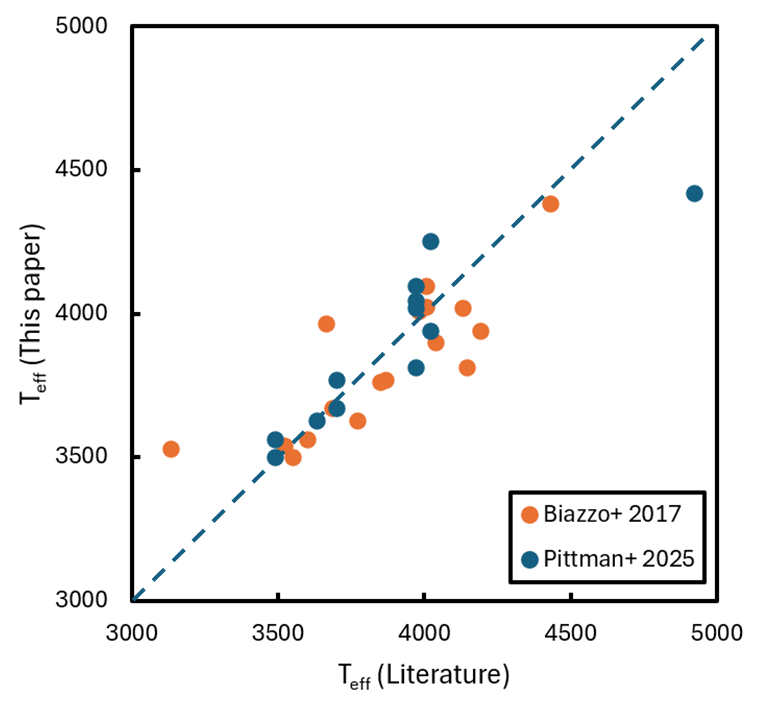}
    \caption{Comparison between the effective temperatures obtained in this paper and Paper I with those from literature. Open orange squares are with \citet{Biazzo2017}; filled blue circles are with \citet{Pittman2025}. The dashed line represents equality.}
    \label{fig:teff_comparison}
\end{figure}

We compared the parameters derived in this paper and in Paper I with values from the literature. In particular, stellar effective temperatures can also be derived from spectra. We found offsets and individual differences that depend on the study considered. The total sample of this study and paper I has 13 stars in common with \citet{Flores2022}. They provide temperatures derived from optical spectra (compiled from the literature) and near-infrared spectra (from their own analysis). Our temperatures are higher by $104\pm 64$~K (r.m.s. scatter of 230~K) for the optical estimates and $254\pm 54$~K (r.m.s. scatter of 193~K) for the near-infrared estimates. In contrast, our temperatures are systematically lower than those of \citet{Gangi2022} by $124\pm 114$~K (r.m.s. scatter of 411~K, 13 stars in common). Our results are instead consistent with those of \citet{Biazzo2017} and \citet{Pittman2025}, with offsets (this work--others) of $-19\pm45$~K (r.m.s. scatter of 179~K; 16 stars in common) and $-11\pm53$~K (r.m.s. scatter of 183~K; 12 stars in common), respectively (see Fig.~\ref{fig:teff_comparison}). Although our temperature estimates are not necessarily more accurate than those from spectroscopy, they are derived homogeneously across the entire sample. Adopting temperatures higher by 100~K would reduce the stellar radii by $\sim5\%$ and increase the inferred ages by $\sim15\%$. 

\subsubsection{Accretion rates}

\begin{figure}[htb!]
    \centering
    \includegraphics[width=0.8\columnwidth]{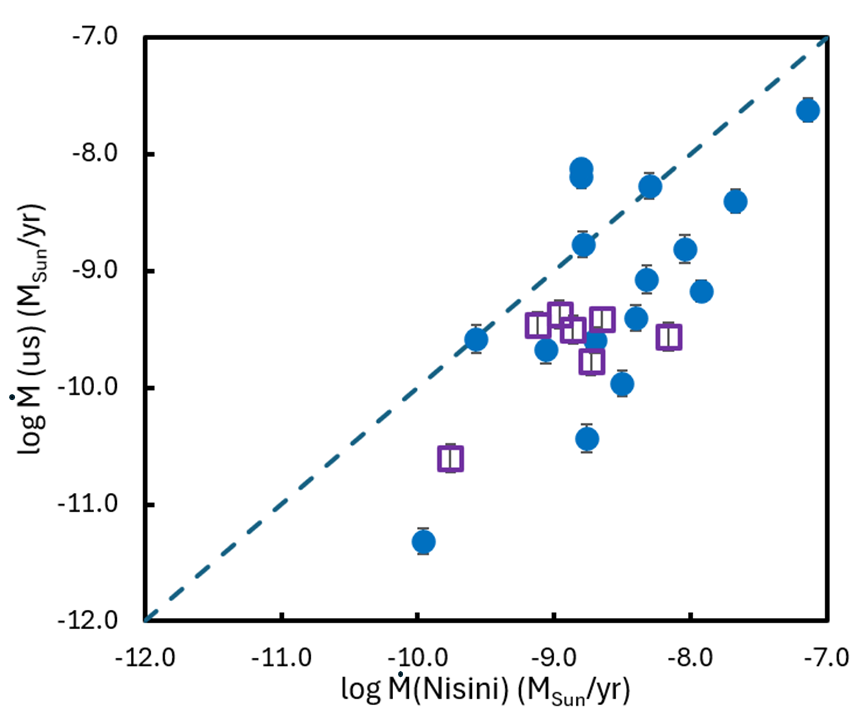}
    \caption{Comparison between the accretion rates obtained in this paper and Paper I with those by \citet{Nisini2018}. Purple open squares are results for the discs analysed in this paper, blue filled circles are for those analysed in Paper I. The dashed line represents equality. }
    \label{fig:mdot_comparison}
\end{figure}

We compared the measurements of the accretion rates determined in our study (including those of Paper I) with the values obtained from high-resolution spectroscopic data (mainly from X-Shooter data) by \citet{Nisini2018} (see Fig.~\ref{fig:mdot_comparison}). We found a good correlation but a significant offset ($-0.66\pm 0.15$~dex with a standard deviation of 0.62~dex, 17 stars in common). The agreement is also fairly good with the values estimated by \citet{Pittman2025}, with a mean difference of $-0.62\pm 0.17$~dex, r.m.s. scatter of 0.60~dex over 12 stars in common. We also compared our values with those recently determined by \citet{Delfini2025}  from the {\it Gaia} XP spectra (prism low resolution), which are available for 38 of the stars considered here. Our estimates are lower by $-0.30\pm 0.10$~dex, with a standard deviation of 0.60~dex. We notice that while we determined the accretion rates directly from the accretion luminosities using space UV data, both \citet{Nisini2018} and \citet{Delfini2025} used the calibration of the equivalent width of the emission of H$\alpha$ from \citet{Alcala2017}.

\subsubsection{Stellar ages}

\begin{figure}[htb!]
    \centering
    \includegraphics[width=0.8\columnwidth]{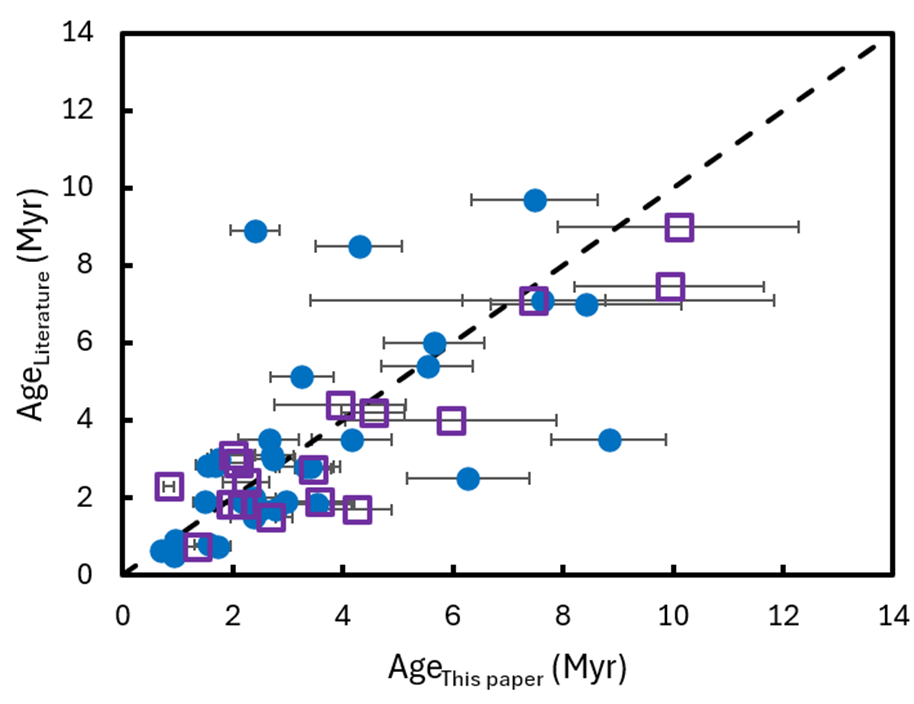}
    \caption{Ages derived in this paper compared to literature values. Purple open squares are results for the discs analysed in this paper, blue filled circles are for those analysed in Paper I.}
    \label{fig:age_comparison}
\end{figure}

Once the best-fitting solution was identified, we derived stellar ages by comparing the measured stellar radii with the \citet{Baraffe2015} isochrones at the corresponding dynamical masses. This approach is well suited to pre-main-sequence stars of approximately solar mass, for which the stellar radius evolves rapidly while the effective temperature remains relatively constant. The uncertainties reported in Table~\ref{tab:derived_data} account only for the internal uncertainties of our analysis. 
Several studies have shown that standard stellar-evolution models that neglect magnetic effects do not fully reproduce pre-main-sequence stellar evolution. However, stellar radii are less sensitive to these effects than effective temperatures. Repeating the age determination using models that include surface spots and magnetically inhibited convection (\citealt{Somers2015,Feiden2016,Towner2025}) yields ages that are, on average, 22\% older. We nevertheless adopt the ages derived from the \citet{Baraffe2015} isochrones to ensure consistency with Paper~I.

Figure \ref{fig:age_comparison} shows a comparison of the ages derived in this paper and in Paper I with other determinations from the literature (see Table \ref{tab:derived_data} for references). Restricting the comparison to the stars analysed in this paper, we found our average ages to be older by $12\pm10$\%, with individual stars showing a standard deviation of 42\%.  We consider this result satisfactory. 

\subsubsection{Disc masses}

\begin{figure}[htb!]
    \centering
    \includegraphics[width=0.8\columnwidth]{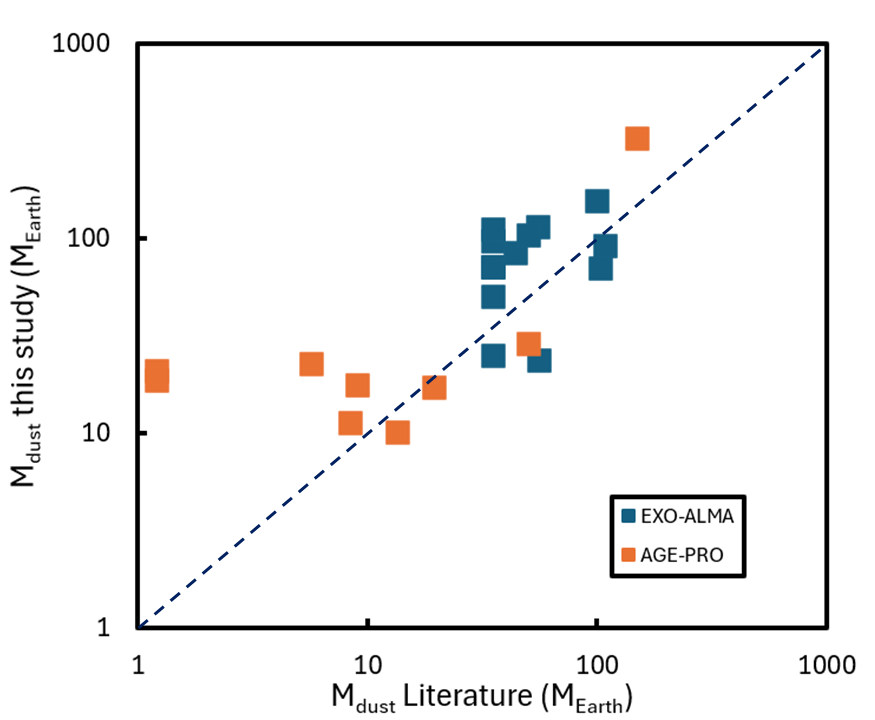}
    \caption{Comparison between the total dust masses obtained in this paper and paper I with those from the {\tt EXO-ALMA} (\citealt{Curone2025}: open blue circles) and {\tt AGE-PRO} (\citealt{Agurto-Gangas2025, Deng2025}: open orange squares) programmes. The dashed line represents equality.}
    \label{fig:dust_mass_comparison}
\end{figure}

The total mass in dust in the discs (almost all in the outer disc) scales directly with the total flux in the ALMA continuum images, with only a weak dependence on disc temperature and grain opacity. Figure \ref{fig:dust_mass_comparison} compares the total dust masses we derive here and in Paper I with those from the {\tt EXO-ALMA} (\citealt{Curone2025}) and {\tt AGE-PRO} (\citealt{Agurto-Gangas2025, Deng2025}) programmes. The agreement is generally good, except for the two lowest-dust-mass stars, Sz 77 and 2MASS J16202863--2442087, for which we obtain higher values.  Both of these have compact, barely resolved discs with noisy density profiles in their outer regions, making our results sensitive to the precise zero point of the ALMA images.

\section{Discussion}
\label{sect:discussion}

\subsection{Evolution of discs}

\begin{figure*}
\sidecaption
    \includegraphics[width=12cm]{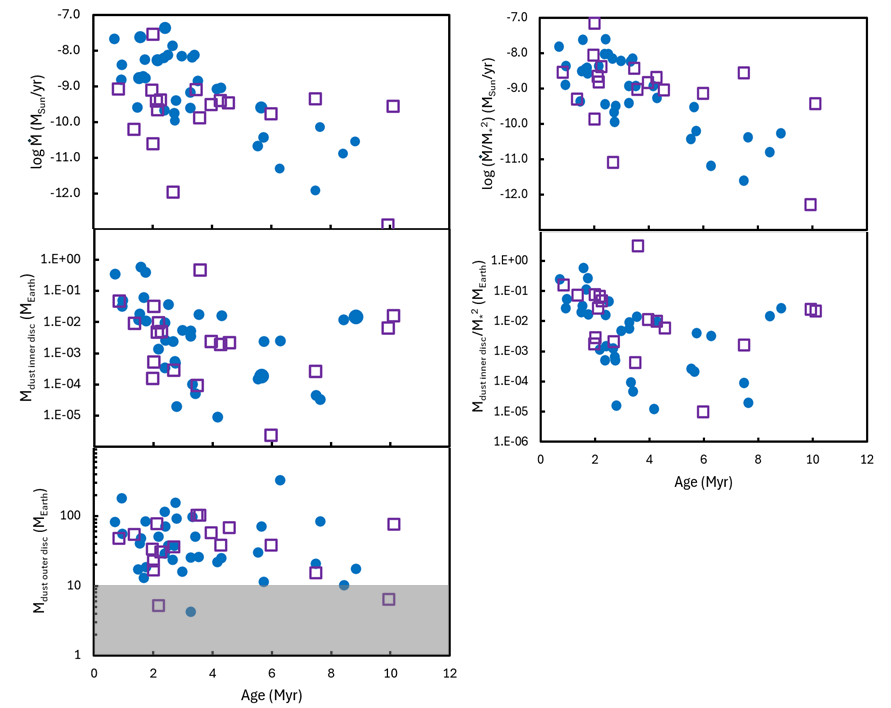}
    \caption{Trends with age for the mass accretion rate (top panel), inner disc dust mass (central panel), and outer disc dust mass (lower panel). Purple open squares show the discs analysed in this paper; blue filled circles show those analysed in Paper I. The grey shaded area in the bottom row marks the region of very low completeness. The left column shows the observed values, while the right column shows values divided by $(M_\star/{\rm M}_\odot)^2$ to account for the dependence of mass accretion rate and inner disc dust mass on stellar mass. }
    \label{fig:age_dependence}
\end{figure*}

Figure \ref{fig:age_dependence} shows how the mass accretion rate, the inner disc dust mass, and the outer disc dust mass change with age. For this purpose, we combined results from this paper and Paper I. The trends in the first two panels are quite clear and confirm previously known results \citep{Hartmann1998, Hernandez2007, Fedele2010}. However, both the mass accretion rates and the inner disc dust masses are in general lower for the low-mass stars studied here than for the solar-type stars in Paper I. We quantified these difference with multiple linear regressions\footnote{We notice that $M_\star$ and age are essentially uncorrelated in our sample the correlation coefficient being -0.004 over 51 stars}. The correlation of \mdot and $M_{\rm dust~inner~disc}$ on stellar mass $M_\star$ and age, are described by the following formulas:
\begin{equation}
\begin{split}
\log{\dot{M}}& = -9.36 + (2.5\pm 0.7)\, (\log{M_\star}+0.14)\, - \\
&(2.2\pm 0.4)\, (\log{\rm Age}-0.47),~{\rm r.m.s.=0.70~dex}
\end{split}
\end{equation}
where \mdot is in \Msun~$yr^{-1}$, and:
\begin{equation}
\begin{split}
\log{M_{\rm dust~inner~disc}} &= -2.57 + (0.5\pm 0.9)\, (\log{M_\star}+0.14)\, - \\
&(1.9\pm 0.6)\, (\log{\rm Age}-0.47),~{\rm r.m.s.=0.84~dex}
\end{split}
\end{equation}
where $M_{\rm dust~inner~disc}$ is in \MEarth, $M_\star$ in units of \Msun and the age in Myr. The significance of the coefficients for $\log{M_{\rm dust~inner~disc}}$ is significantly enhanced if we only consider stars with age younger than 8~Myr and neglect Haro 6-33, which appears to be discrepant. In this case, we obtain:
\begin{equation}
\begin{split}
\log{M_{\rm dust~inner}} &= -2.67 + (1.0\pm 0.7)\, (\log{M_\star}+0.13)\, - \\
&(3.2\pm 0.5)\, (\log{\rm Age}-0.43),~{\rm r.m.s.=0.83~dex}.
\end{split}
\end{equation}
The slope of the accretion rate's dependence on stellar mass agrees, within the errors, with the $M_\star^2$ relation adopted in previous studies \citep[e.g.][]{Manara2023}, as shown in the right panels of Fig.~\ref{fig:age_dependence}. The shallower dependence on the age for $M_\star$ and the steeper one found for $M_{\rm dust~inner~disc}$ may be explained by how the fraction of stars showing dust filtration varies with $M_\star$, as discussed in the next subsection.

Regarding the mass in the outer disc, we noted that the stars in this study are biassed towards those with prominent discs \citep{vanderMarel2023}. Although a large fraction of very young stars have prominent discs, fewer than a third of those older than 5~Myr host a clearly detectable disc (\citealt{Rigliaco2025}; see also discussion in Sect.~\ref{sect:bias}). Once this bias is taken into account, a trend between outer disc dust mass and age is also present, though less clear than for the inner disc.

Overall, these data indicate that this sample is dominated by young or transitional discs. Classical transitional discs are defined as protoplanetary discs with little or no near-IR excess ($\lambda<10$~$\mu$m) and a significant excess comparable to the median of Taurus \citep{Furlan2006} at longer wavelengths \citep{Strom1989, Calvet2002, Calvet2005}.

\begin{figure}
    \centering
    \includegraphics[width=0.8\columnwidth]{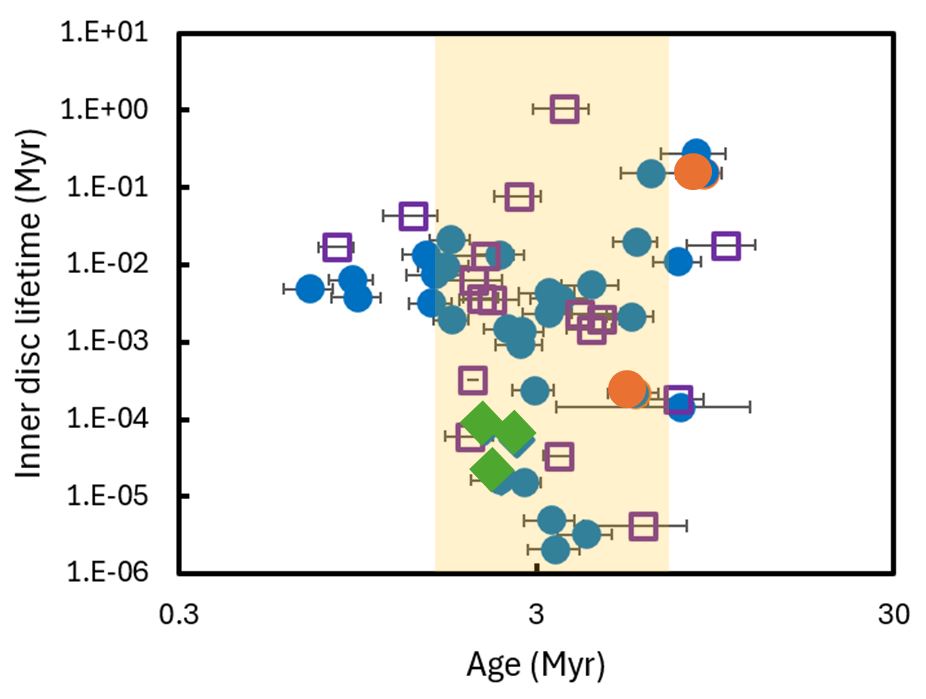}
    \includegraphics[width=0.8\columnwidth]{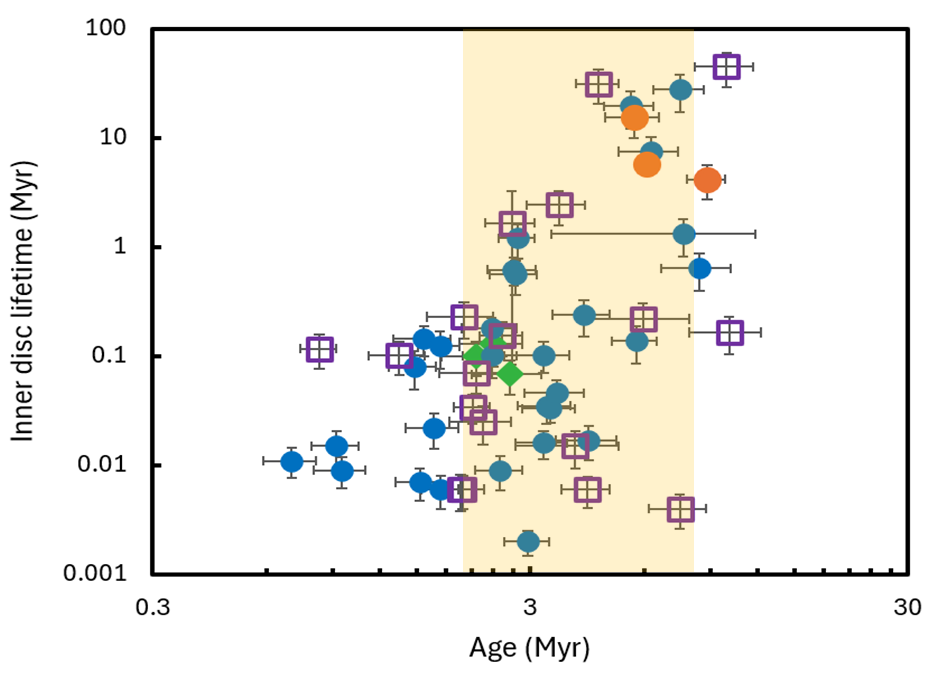}
    \caption{Trend of inner disc lifetime with age. Upper panel: gas masses obtained by assuming a constant gas-to-dust ratio of 100. Lower panel: gas masses determined directly from our model. Purple open squares show the discs analysed in this paper; blue filled circles show those analysed in Paper I. Green diamonds mark discs with large asymmetries, and filled orange circles mark stars hosting accreting planets. The shaded region indicates the 2–8~Myr epoch, when giant planet cores are likely still forming. }
    \label{fig:disc_lifetime}
\end{figure}

Assuming a constant gas-to-dust mass ratio (100 in our model), we estimate the inner-disc lifetime by dividing the inner-disc mass, defined here as the mass within the snow line, by the stellar mass accretion rate. Under this approximation, this timescale is expected to reflect the basic properties of viscously evolving discs \citep{Lynden-Bell1974}. For systems younger than 2~Myr, the inferred inner-disc lifetime is comparable to the viscous timescale for a viscosity parameter of $\sim0.001$ \citep{Manara2016,Lodato2017}, corresponding to a timescale of approximately $10^4$~yr (see the upper panel of Fig.~\ref{fig:disc_lifetime}). 

However, for 17 of the 35 discs around stars aged 2--8~Myr, the inner-disc lifetime estimated in this way is $<1000$~yr. In several cases, the inferred timescales are sufficiently short to raise concerns about this interpretation. We emphasise that our model does not directly constrain the total inner-disc mass, but only its dust mass; the total mass is inferred by adopting a fixed gas-to-dust ratio. The combination of relatively high accretion rates and low inferred dust masses could instead indicate that these discs are depleted in large grains, resulting in an effectively higher gas-to-dust ratio.

Indeed, this discrepancy is largely removed when using the inner-disc masses derived directly from the model, as shown in the lower panel of Fig.~\ref{fig:disc_lifetime}. In this case, the inferred inner-disc lifetime scales approximately with stellar age, providing a more plausible explanation for the persistence of these discs. Long inner-disc lifetimes could in principle be produced by very low viscosities (of the order of $10^{-5}$). However, low viscosities favour grain growth \citep{Birnstiel2012,Zagaria2022}, which is inconsistent with the low values of $R_{\rm max,grain,inner}$ inferred for some of these discs and would therefore require an alternative explanation. Because the gas mass is not directly constrained by our observations, the inferred gas-to-dust ratios are subject to substantial uncertainties. We therefore interpret this result only as qualitative evidence that some inner discs may have significantly elevated gas-to-dust ratios. Furthermore, evolved discs may contain a substantial fraction of their mass in planetesimals, which do not contribute significantly to the observed continuum emission.

\subsection{Relation to planet formation}

\begin{figure}[htb!]
    \centering
    \includegraphics[width=0.8\columnwidth]{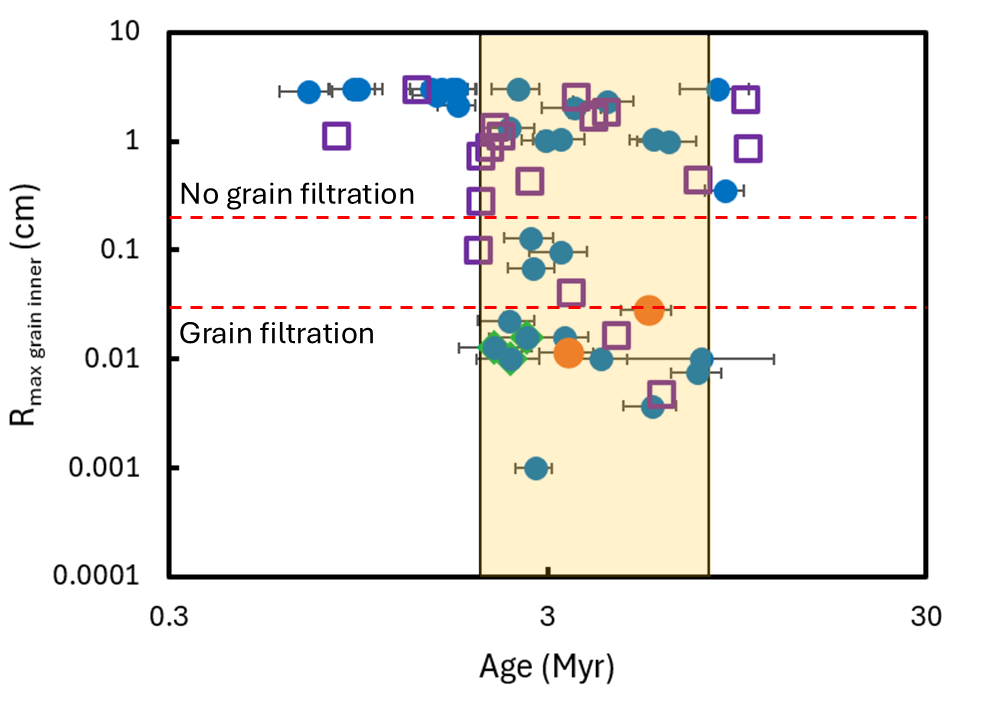}
    \caption{Trend of the maximum radius of the inner disc grains with age. Purple open squares are results for the discs analysed in this paper, blue filled circles are for those analysed in Paper I. Green diamonds are discs exhibiting large asymmetries. Filled orange circles are stars hosting accreting planets. The shaded area refers to the epoch between 2--8~Myr where giant planet cores are likely building-up.  We considered as discs with evidence of dust filtration those with $R_{\rm max,grain,inner}<0.03$ cm (see \citealt{Weber2018}) and with no filtration those with $R_{\rm max,grain,inner}>0.2$~cm.}
    \label{fig:age_rmaxgraininner_2}
\end{figure}

\begin{figure}[htb!]
    \centering
    \includegraphics[width=0.8\columnwidth]{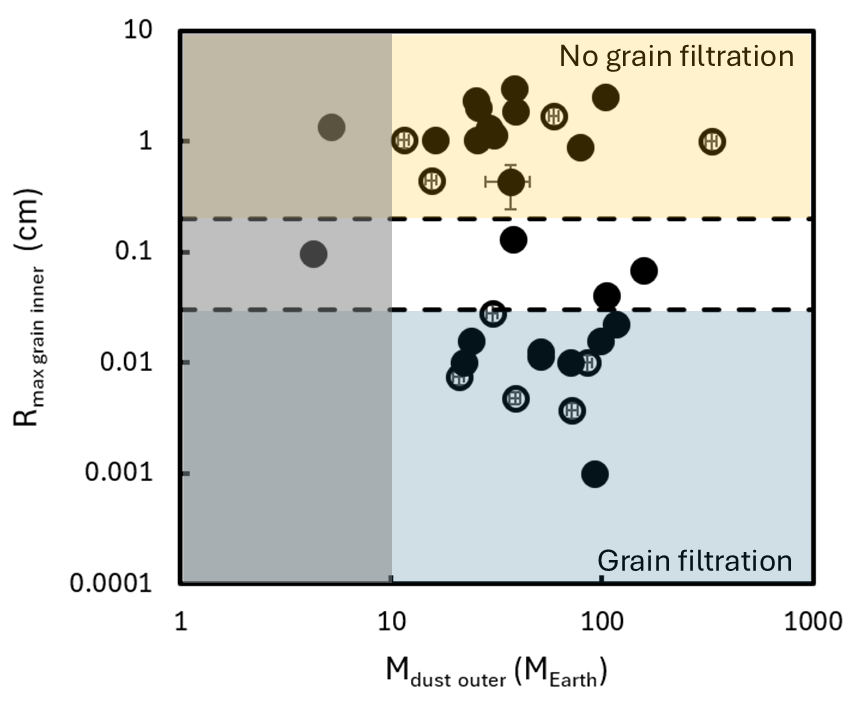}
    \caption{Trend of the maximum radius of the inner disc grains with dust mass in the outer disc. Filled symbols are stars with $2<$Age$<5$~Myr, open symbols are stars with $5<$Age$<8$~Myr. We considered as discs with evidence of dust filtration those with $R_{\rm max,grain,inner}<0.03$ cm (see \citealt{Weber2018}) and with no filtration those with $R_{\rm max,grain,inner}>0.2$~cm. The grey shadowed region at low values of the dust mass in the outer disc signals high incompleteness in our study.}
    \label{fig:mdust_rmaxgraininner}
\end{figure}

\begin{figure}[htb!]
    \centering
    \includegraphics[width=0.8\columnwidth]{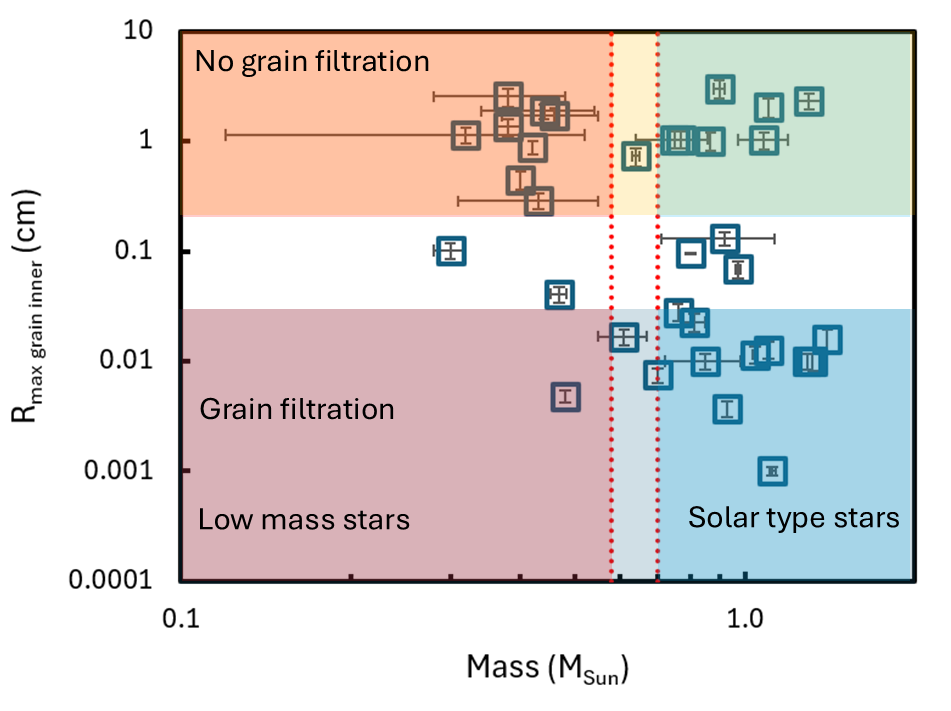}
    \caption{Maximum grain radius in the inner disc as a function of stellar mass for stars with ages of 2--8~Myr, during which giant-planet cores are expected to form. We classify discs as showing evidence of dust filtration if $R_{\rm max,grain,inner}<0.03$~cm (see \citealt{Weber2018}), and as showing no evidence of filtration if $R_{\rm max,grain,inner}>0.2$~cm. Solar-type stars are defined as those with $0.7<M_\star<1.5$~\Msun, while low-mass stars considered progenitors of main-sequence early M dwarfs have $0.30<M_\star<0.58$~\Msun.}
    \label{fig:dust_filtering}
\end{figure}

\begin{figure}[htb!]
    \centering
    \includegraphics[width=0.8\columnwidth]{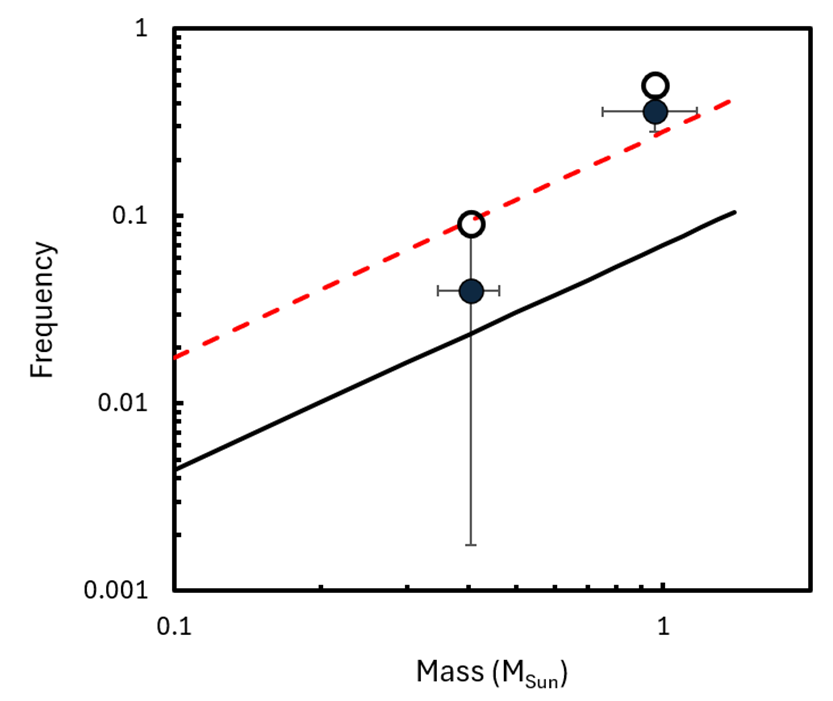}
    \caption{Frequency of discs showing evidence of dust filtration in the age range 2--8 Myr as a function of the mass of the star from Paper I and this study (black empty circles) and corrected for the bias related to the mass of the discs discussed in Sect. \ref{sect:data}(black solid circles). The solid line is the frequency of the giant planets as a function of stellar mass from \citet{Johnson2010}. The dashed line is obtained considering the same slope but assuming that the frequency of giant planets at 1.0 \Msun is 0.27 rather than 0.07. }
    \label{fig:frequency}
\end{figure}

Figure~\ref{fig:age_rmaxgraininner_2} shows the maximum grain radius in the inner disc as a function of stellar age. As found in Paper~I, large grains are relatively scarce in the inner discs of several stars aged 2--8~Myr compared with younger systems. A similar trend is observed in systems hosting known accreting planets. Adopting $R_{\rm max,grain,inner}<0.03$~cm as a criterion for large-grain depletion \citep[see also][]{Weber2018}, we find that 14 of the 36 stars in this age range show evidence of dust filtration. Such depletion may result from barriers that prevent the inward drift of large grains \citep{Zhang2018}, although alternative interpretations remain possible \citep[e.g.][]{Zhang2015}.

If large-grain depletion is interpreted as a signature of a massive core undergoing giant-planet formation, even in the absence of a direct planetary detection \citep{Zhang2018}, our results suggest that the formation of such cores ($>$~\MEarth, see Sect.~\ref{sect:Introduction}) requires at least $\sim2$~Myr in the systems considered here. Dust filtration therefore appears to occur in a substantial fraction of discs during the 2--8~Myr interval.
Interestingly, the four stars older than 8~Myr in our sample also retain large grains in their inner discs, despite having very low inner-disc dust masses ($<0.01$~M$_{\rm Earth}$). This finding is consistent with \citet{Liu2026}, who reported systematically larger mean grain sizes in the inner discs of Upper Scorpius (mean age $\sim9$~Myr) than in the younger Taurus population. Together, these results suggest that the depletion of large grains is not simply a consequence of disc evolution.
However, the efficiency of this process depends on grain evolution, as coagulation can replenish the population of large grains in the inner disc even when filtration is operating \citep{Zhu2012, Pfeil2025}. Consequently, the presence of large grains in the inner disc does not necessarily exclude filtration, while their scarcity may trace particularly efficient filtration. Our criterion should therefore be interpreted as identifying discs with evidence of strong filtration, rather than providing a complete census of its occurrence. The fraction of discs satisfying this criterion may thus underestimate the fraction in which filtration operates, a caveat that should be considered when interpreting the implications of our results for giant-planet formation.

The occurrence of dust filtration is also correlated with the outer-disc dust mass (Fig.~\ref{fig:mdust_rmaxgraininner}). Discs showing evidence of filtration generally have more massive outer dust reservoirs than the remaining systems, with median dust masses of $60\pm8$~\MEarth and $31\pm16$~\MEarth, respectively. Notably, all discs showing evidence of filtration have outer-disc dust masses $>20$~\MEarth. This trend is consistent with a scenario in which the massive cores that eventually form giant planets preferentially develop in more massive discs. It also suggests that the lower-mass discs excluded by the selection criteria described in Sect.~\ref{sect:data}, which typically have dust masses $<10$~\MEarth (grey region in Fig.~\ref{fig:mdust_rmaxgraininner}), are unlikely to undergo substantial dust filtration.

We further find evidence for a dependence of the dust-filtration frequency on stellar mass (Fig.~\ref{fig:dust_filtering}). Restricting the analysis to stars aged 2--8~Myr and excluding uncertain cases ($0.03<R_{\rm max,grain,inner}<0.2$~cm, white region in Fig.~\ref{fig:dust_filtering}), we compared solar-type progenitors ($0.7<M_\star<1.4$~\Msun) with early M-type progenitors ($0.30<M_\star<0.58$~\Msun; \citealt{Pecaut2013}). Dust filtration is identified in $11/22$ ($50\pm11\%$) of the solar-type systems, with a mean stellar mass of $0.96\pm0.21$~\Msun, compared to $1/11$ ($9\pm9\%$) of the early M-type systems, with a mean mass of $0.40\pm0.06$~\Msun. A Fisher's exact test gives a probability of 2.2\% of obtaining such a difference by chance, providing evidence for a dependence of the dust-filtration frequency on stellar mass. If dust filtration traces the formation of massive cores that subsequently grow into giant planets, this trend is qualitatively consistent with the lower occurrence of giant planets around M-type stars. It does not, however, reflect the higher occurrence of lower-mass planets around low-mass stars.

To place this result in the context of giant-planet occurrence, Fig.~\ref{fig:dust_filtering} compares the frequency of dust filtration among 2--8~Myr-old systems with the stellar-mass dependence of giant-planet occurrence measured for field stars by \citet{Johnson2010}. Combining the samples from Paper~I and this study, the ratio of filtration frequencies between solar-type and early M-type stars is consistent with the relation $f=0.07~(M_\star/M_\odot)^{1.2}$ derived by \citet{Johnson2010}. The absolute filtration frequencies, however, are substantially higher. Part of this discrepancy may arise from the selection against low-mass discs discussed in Sect.~\ref{sect:data}. Because such discs are unlikely to undergo significant dust filtration, we correct the observed frequencies for the fraction of discs with 0.89-mm fluxes below 25~mJy that are excluded from our sample. The resulting corrected frequencies are shown by the filled symbols in Fig.~\ref{fig:dust_filtering}. Although the correction lowers the measured frequencies, it does not remove the discrepancy with the giant-planet occurrence rate measured for field stars. At $1.0$~\Msun, the corrected filtration frequency is $0.28\pm0.07$, compared with $\sim0.07$ predicted by the relation of \citet{Johnson2010}.

This remaining difference may partly reflect the distinct star-forming environments represented by the two samples. Our sample was selected to include stars with long-lived discs in nearby star forming regions and it is likely not representative of the progenitors of the old field stars considered by \citet{Johnson2010}. Most field stars are expected to have formed in substantially more massive associations than the relatively sparse regions currently found in the solar neighbourhood, such as Taurus, Lupus, and Upper Scorpius \citep[e.g.][]{Lada2003}. Moreover, \citet{Gratton2024,Gratton2025} found that the frequency of giant planets depends on the mass and age of the stellar association in which the host stars formed, considering only single stars and binaries with sufficiently distant companions. To quantify the possible effect of the star-forming environment, we note that most of the 2--8~Myr-old stars in our sample belong to Taurus (15 stars) or Lupus (12 stars), with most of the remainder associated with similarly sparse regions, including CrA-North and Chamaeleon. Integrating the stellar mass function of Taurus derived by \citet{Esplin2019} gives a total stellar mass of $\sim130$~\Msun, increasing to $\sim170$~\Msun when lower-mass companions are included. Lupus contains approximately half as many stars as Taurus \citep{Luhman2020,Luhman2023}. These values are likely lower limits, as star formation is still ongoing in both regions. If their stellar masses approximately double by the end of star formation, Equation~(3) of \citet{Gratton2025} predicts a giant-planet occurrence rate of $0.3\pm0.1$ for solar-type stars formed in associations of this mass. This is comparable to the corrected filtration frequency measured in our sample.

This agreement suggests that the higher filtration frequency measured in our sample compared with the giant-planet occurrence rate inferred from the field-star population may, at least in part, reflect differences in the star-forming environments represented by the two samples. In this interpretation, the connection between dust filtration and giant-planet formation may depend not only on stellar and disc properties, but also on the environment in which planetary systems form.

Finally, the only low-mass star in our sample showing evidence of dust filtration is Sz~91, consistent with the results found by \citet{Mauco2021}. Its dust disc is characterised by a ring that peaks at $\sim90$~au and has an inner radius of $\sim65$~au \citep{Tsukagoshi2014}. Intriguingly, Maio et al. (in prep.) have reportedly obtained evidence for a giant planet within the cavity using high-contrast imaging with ERIS at the VLT. If confirmed, such a detection would provide independent evidence that the dust-filtration signature observed in Sz~91 is associated with giant-planet formation. 
A similar behaviour linking the presence of a disc cavity to the possible presence of dust filtration was also noted in Section 4.2 of Paper I, for the whole sample of solar-type stars, where a correlation between the disc classification of \citet{Parker2022} and the values of $R_{\rm max, grain}$ (or the slope in the grain distribution in the inner disc) was found.

\section{Conclusions}
\label{sect:conclusions}

In a previous paper \citep{Gratton2026}, we used a simple model to simultaneously reproduce the SEDs and high-resolution ALMA dust-continuum images of 33 discs around solar-type stars ($0.7<M_\star<1.5$~\Msun). The stars were selected to have dynamical masses derived from the kinematics of their molecular gas discs, which allowed us to estimate relatively accurate individual ages from the age--radius relation. The selected sample therefore enabled us to investigate the evolution of several disc properties as a function of age. In particular, we found that approximately half of the discs with ages of 2--8~Myr show a depletion of large grains in their inner discs ($R_{\rm max,grain,inner}<0.03$~cm) whereas no such depletion is observed in discs younger than 2~Myr. We interpreted this result as possibly arising from the filtration of large grains by pressure structures associated with the growth of the cores of forming giant planets to masses of a few tens of Earth masses. If this interpretation is correct, the frequency of discs showing evidence of dust filtration should decrease towards lower stellar masses, given the well-established lower occurrence rate of giant planets around evolved low-mass stars.

We then applied the same methodology described in Paper I to a sample of 18 low-mass stars with prominent discs, selected according to the same criteria adopted in Paper I, thereby extending our analysis to a wider range of stellar masses. The results confirm that there is no evidence for grain filtration in very young discs (ages $<2$~Myr). Among older systems in the 2--8~Myr age range, we found evidence of dust filtration in only one of the 11 discs around stars with $0.30<M_\star<0.58$~\Msun, which likely are progenitors of early M-type stars of the main-sequence. The resulting fraction of stars showing evidence of filtration, $0.09\pm0.09$, is therefore significantly lower than that measured for stars with masses similar to the Sun, $0.50\pm0.10$. The probability that this difference arises by chance is only 1.9\%, suggesting that the dependence of the filtration frequency on stellar mass is likely significant. This result supports the hypothesis that dust filtration is associated with the growth of the cores of forming giant planets, whose occurrence is known to decrease towards lower stellar masses. We also find that discs showing evidence of filtration are generally relatively massive, with outer-disc dust masses $>20$~\MEarth, suggesting that efficient grain filtration preferentially occurs in massive discs. The absence of a corresponding increase in filtration frequency around low-mass stars, despite the higher occurrence of Neptune-like and lower-mass planets in these systems \citep{Mulders2015,Mulders2018}, suggests that grain filtration may not be an efficient mechanism associated with the formation of these lower-mass planets.

However, the frequencies of discs showing evidence of dust filtration in the 2--8~Myr age range measured in Paper~I and in this study are higher than the observed frequency of giant planets around field stars of comparable mass. Part of this difference may result from the bias towards massive, prominent discs inherent to the samples considered in both studies, although this selection effect does not appear to fully account for the discrepancy. An additional possibility is that the difference reflects the distinct star-forming environments represented by the two samples. Previous studies have suggested that the occurrence rate of giant planets may depend on the properties of the stellar associations in which their host stars formed, with higher frequencies inferred for some young, relatively sparse associations than for the older field population. If dust filtration is associated with giant-planet formation, differences in the formation environments could therefore contribute to the higher filtration frequency observed in our sample. This possibility could be tested by extending the analysis to discs in more massive star-forming regions, such as the Orion Nebula Cluster, using high-resolution ($<50$~mas) observations (see, e.g., \citealt{Otter2021}).

\begin{acknowledgements}
This paper makes use of the following ALMA data: ADS/JAO.ALMA\#2015.1.01083.S. ALMA is a partnership of ESO (representing its member states), NSF (USA) and NINS (Japan), together with NRC (Canada), NSTC and ASIAA (Taiwan), and KASI (Republic of Korea), in cooperation with the Republic of Chile. The Joint ALMA Observatory is operated by ESO, AUI/NRAO and NAOJ.
R.G. and E.R. gratefully acknowledges support from the “Programma di Ricerca Fondamentale INAF 2023” of the Italian National Institute of Astrophysics (INAF Large Grant 2023 “NextSTEPS”). 
\end{acknowledgements}
\bibliographystyle{aa} 
\bibliography{biblio} 

\begin{appendix}

\section{Input data }
\label{app:input_data}
Tables \ref{tab:uv_data}, \ref{tab:input_data} and \ref{tab:alma_data} give the input data used in our analysis. Tables \ref{tab:output_data} and \ref{tab:derived_data} list the output parameters. In the following, we give some comments about individual objects.

CW Tau: Following our standard procedure, the K3 spectral type adopted for CW Tau \citep{Luhman2010} corresponds to an intrinsic colour of $(B_p-R_p)_0=1.21$ according to the tables of \citet{Pecaut2013}. Comparing this value with the observed colour, $B_p-R_p=2.139$, yields an extinction of $A_V=1.956$~mag. Using these values, we derive an age of $8.5\pm1.9$~Myr, substantially older than the median age of the L1495 group (1.8~Myr; Gratton et al., in prep.). This age also appears inconsistent with other properties of CW Tau, including its relatively high accretion rate ($\log{\dot{M}}=-7.57\pm0.07$~\Msun~yr$^{-1}$) and the presence of prominent bipolar jets \citep{McGroarty2004}.
The anomalously old age is likely related to the contribution of accretion-related emission at optical wavelengths, which can dominate the stellar flux at wavelengths shorter than $\sim0.7$~$\mu$m. Indeed, high veiling, with values of approximately unity at optical wavelengths, has been reported for CW Tau by \citet{Fischer2011} and \citet{Herczeg2014}. If the optical emission is substantially contaminated by accretion, the stellar photospheric luminosity—and hence the stellar radius and age—cannot be reliably constrained from the optical fluxes. We therefore adopt the extinction of $A_V=1.74$~mag obtained by \citet{Herczeg2014} and assume an age of $\sim2$~Myr, consistent with the L1495 group. These assumptions imply a stellar radius of $R_*\sim1.95$~\RSun.

DM Tau:	\citet{Longarini2025} gives a mass of $0.47\pm 0.02$ \Msun that we adopt in our analysis. \citet{Willson2016} modelled their sparse aperture masking observations with an additional point source at 43 mas ($\sim 6$\, au) with an absolute magnitude of $M_K = 11.0\pm 0.3$\, mag and a significance of 4.27 $\sigma$. If this point source is a physical companion, it would have a planetary mass. Binarity is possibly supported by a moderately high value of the Renormalised Unit Weight Error (RUWE) parameter from {\it Gaia} (RUWE=1.487). However, these evidences for a companion may result from an asymmetric light distribution in the inner disc as proposed by several authors \citep{Hashimoto2021, Terada2023, Liu2024}. 

Haro 6-33, alias IRAS 04385+2550: This star is associated with powerful jets \citep{Anathpindika2008, Bally2012}, which are indicative of active and relatively young stellar evolution. However, assuming a single-star configuration, we derive an age of $6.7\pm1.4$~Myr. This relatively old age is inconsistent with the presence of powerful jets. The discrepancy may be alleviated if Haro 6-33 is instead a binary system, as suggested by its high Gaia RUWE value (5.057). Following the approach adopted in \citet{Gratton2024}, we found that the RUWE signal is best reproduced by a system consisting of a primary with a mass of 0.38~\Msun\ and a secondary of 0.22~\Msun\ at a separation of 1.7~au. The resulting total mass of 0.60~\Msun\ is consistent with the value of $0.60\pm0.10~$\Msun derived by \citet{Simon2019} from the kinematics of the disc. Adopting 0.38~\Msun\ for the primary mass yields an age of $3.6\pm0.6$~Myr, which is the value adopted in our analysis.

Sz 73: also in this case we derive an old age related to a high absorption. The spectral type of K7 is from \citet{Alcala2014}, which also obtained a high reddening and a similar old age for this star. We did not find reason to modify this analysis.

\FloatBarrier
\begin{table*}[h]
\caption{Space UV data.}
\centering
\tiny
%\begin{adjustbox}{angle=-90}
\begin{tabular}{lrrccccccc}
\hline
\hline
Star & RA	&	Dec	& \multicolumn{2}{c}{HST}  &\multicolumn{2}{c}{Galex} & \multicolumn{3}{c}{XMM/UVOT} \\	
&(degree)	&	(degree)	&		COS	&	STIS 	&	FUV	&	NUV	 &	UVW2	&	UVM &	UVW1 \\
\hline
CW Tau	&	63.57089	&	28.18261	&	Yes	&	&	Yes	&	Yes	&	Yes	&	Yes	&	Yes \\
CX Tau	&	63.69947	&	26.80296	&		&		&	Yes	&		&		&		&	\\
CY Tau	&	64.39058	&	28.34622	&	Yes	&	Yes	&		&		&		&		&	\\
FT Tau	&	65.91332	&	24.93719	&		&		&		&		&		&		&	\\
DM Tau	&	68.45311	&	18.16936	&	Yes	&		&		&		&		&		&	\\
HO Tau	&	68.83426	&	22.53730	&		&		&		&		&		&		&	\\
DO Tau	&	69.61915	&	26.18031	&		&		&		&		&		&		&		\\
Haro 6-33	&	70.41180	&	25.94068	&		&		&		&		&		&		&	\\
HM Lup	&	236.96088	&	-35.47660	&	Yes	&		&		&		&		&		&	\\
Sz 73	&	236.98719	&	-35.24310	&		&		&		&		&		&		&	\\
2MASS J16000236-4222145	&	240.00976	&	-42.37083	&		&		&		&		&		&		&	\\
Sz 91	&	241.79822	&	-39.06329	&		&		&		&		&		&		&	\\
2MASS J16085324-3914401	&	242.22178	&	-39.24459	&		&		&		&		&		&		&	\\
Sz 111	&	242.22780	&	-39.62875	&	Yes	&		&		&		&		&		&	\\
Sz 114	&	242.25765	&	-39.08689	&	Yes	&		&		&		&		&		&	\\
Sz 123	&	242.71489	&	-38.88726	&		&		&		&		&		&		&	\\
2MASS J16145026-2332397	&	243.70939	&	-23.54447	&		&		&		&			&		&		&	\\
2MASS J16163345-2521505	&	244.13931	&	-25.36417	&		&		&		&		&		&		&	\\
\hline
\end{tabular}
%\end{adjustbox}
\label{tab:uv_data}
\end{table*}
\FloatBarrier

\FloatBarrier
\begin{table*}[h!]
\caption{Input parameters.}
\centering
\tiny
\begin{tabular}{lcccccccccc}
\hline
\hline
Star&$\pi$&G&$B_P-R_P$&Sp.T.&$A_V$&$M_\star$&ref&$i$&$PA$&ref\\
&(mas)&(mag)&(mag)&&(mag)&(\Msun)&&(deg)&(deg)&\\
\hline
CW Tau	&	7.602	&	12.775	&	2.139	&	K3	&	1.74	&	0.64	$\pm$	0.01	&	2	&	58.5	&	61.4	&	9	\\
CX Tau	&	7.891	&	12.601	&	2.406	&	M1.5	&	0.581	&	0.38	$\pm$	0.02	&	2	&	55.1	&	66.2	&	5	\\
CY Tau	&	7.916	&	12.480	&	2.211	&	M1.5	&	0.171	&	0.30	$\pm$	0.02	&	2	&	28.5	&	153.0	&	3	\\
FT Tau	&	7.681	&	13.604	&	2.569	&	M2.8	&	0.259	&	0.35	$\pm$	0.05	&	1	&	35.5	&	121.8	&	1	\\
DM Tau	&	6.942	&	13.096	&	2.029	&	M2	&	0.000	&	0.47 $\pm$ 0.02			&	13	&	39.8	&	335.7	&	8	\\
HO Tau	&	6.080	&	13.982	&	1.459	&	M0.5	&	0.000	&	0.44	$\pm$	0.10	&	1	&	55.0	&	116.3	&	1	\\
DO Tau	&	7.219	&	12.381	&	2.059	&	M1	&	0.000	&	0.54	$\pm$	0.07	&	1	&	27.6	&	170.0	&	1	\\
Haro 6-33	&	6.266	&	15.969	&	3.201	&	M0.5	&	2.591	&	0.38	&	7	&	61.2	&	158.5	&	12	\\
HM Lup	&	6.381	&	13.498	&	2.413	&	M3	&	0.000	&	0.42			&	6	&	31.5	&	47.89	&	4	\\
Sz 73	&	6.336	&	14.490	&	2.750	&	K7	&	2.211	&	0.86	$\pm$	0.23	&	1	&	29.5	&	110.0	&	1	\\
2MASS J16000236-4222145	&	6.235	&	14.737	&	2.922	&	M4	&	0.000	&	0.37	$\pm$	0.04	&	1	&	65.7	&	160.45	&	1	\\
Sz 91	&	6.274	&	14.377	&	2.705	&	M0.5	&	1.547	&	0.48			&	10	&	48.1	&	18.0	&	11	\\
2MASS J16085324-3914401	&	6.135	&	15.022	&	3.328	&	M3	&	1.744	&	0.43	$\pm$	0.12	&	1	&	48.0	&	110.18	&	4	\\
Sz 111	&	6.315	&	13.306	&	2.133	&	M1	&	0.090	&	0.61	$\pm$	0.06	&	1	&	52.0	&	40.0	&	1	\\
Sz 114	&	6.379	&	13.642	&	2.847	&	M4	&	0.000	&	0.32	$\pm$	0.20	&	1	&	15.8	&	148.73	&	1	\\
Sz 123	&	6.166	&	14.247	&	2.425	&	M1	&	0.705	&	0.46	$\pm$ 0.09		&	1	&	48.0	&	156.0	&	1	\\
2MASS J16145026-2332397	&	6.898	&	14.471	&	1.942	&	M3	&	0.000	&	0.40			&	6	&	10.0	&	136.0	&	6	\\
2MASS J16163345-2521505	&	6.271	&	13.982	&	2.334	&	M0.5	&	0.601	&	0.51			&	6&	62.0	&	242.0	&	6	\\
\hline
\end{tabular}
\tablebib{
(1) \citet{Braun2021};
(2) \citet{Simon2019};
(3) \citet{Guilloteau2011};
(4) \citet{Guerra2025};
(5) \citet{Facchini2019};
(6) \citet{Trapman2025};
(7) This paper;
(8) \citet{Izquierdo2025};
(9) \citet{Ueda2022};
(10) \citet{Tsukagoshi2019};
(11) Maio et al. 2026, in prep.;
(12) \citet{Aizawa2020}.
(13) \citet{Longarini2025}}
\label{tab:input_data}
\end{table*}
\FloatBarrier

\FloatBarrier
\begin{table*}[t]
\caption{ALMA data.}
\centering
\tiny
\begin{tabular}{lcccccccc}
\hline
\hline
Star&Program&{\tt CASA} &Band&Cont. Sens.&Res.&F$_{870}$ & F$_{1330}$&Center\\
&&Version&&(mJy/beam)&(arcsec)&(mJy)&(mJy)&(mJy/beam)\\
\hline
CW Tau	&	2019.1.01108.S	&6.1.1-15&	6	&	0.0143	&	0.060	&	160	&	74	&	25.0	\\
CX Tau	&	2016.1.00715.S	&4.7.2-REL& 6	&	0.0202	&	0.034	&	25	&	9.4	&	16.7	\\
CY Tau	&	2022.1.01284.S	&6.4.1-12&	6	&	0.0162	&	0.048	&	164	&	79	&	11.4	\\
FT Tau	&	2016.1.01164.S	&5.1.1-5&	6	&	0.0785	&	0.111	&	122	&	63	&	91.1	\\
DM Tau	&	2021.1.01123.L	&6.2.1-7&	7	&	0.0541	&	0.072	&	24	&	8.9	&	7.3	\\
HO Tau	&	2016.1.01164.S	&5.1.1-5&	6	&	0.0785	&	0.108	&	44	&	17	&	22.4	\\
DO Tau	&	2019.1.00496.S	&6.2.1-7&	6	&	0.0278	&	0.124	&	253	&	108	&	25.4	\\
Haro 6-33	&	2021.1.00854.S	&6.2.1-7&	6	&	0.0278	&	0.060	&	...	&	38	&	35.3	\\
HM Lup	&	2022.1.00154.S	&6.4.1-12&	6	&	0.0293	&	0.030	&	14	&	5.5	&	17.7	\\
Sz 73	&	2018.1.01458.S	&5.4.0-70&	6	&	0.0551	&	0.043	&	30	&	10.8	&	35.6	\\
2MASS J16000236-4222145	&	2018.1.01054.S	&5.4.0-70&	6	&	0.0330	&	0.021	&	...	&	50	&	21.9	\\
Sz 91	&	2012.1.00761.S	&4.3.1-REL&	7	&	0.0457	&	0.134	&	34	&	9.5	&	-0.1	\\
2MASS J16085324-3914401	&	2022.1.00828.S	&6.5.4-9&	7	&	0.0548	&	0.013	&	20	&	8.0	&	105	\\
Sz 111	&	2018.1.00689.S	&5.4.0-70&	6	&	0.0121	&	0.021	&	...	&	60.3	&	-3	\\
Sz 114	&	2016.1.00484.L	&5.1.1-5&	7	&	0.0191	&	0.034	&	96	&	42	&	16.0	\\
Sz 123	&	2018.1.01458.S	&5.4.0-70&	6	&	0.0692	&	0.044	&	41	&	16	&	3.0	\\
2MASS J16145026-2332397	&	2021.1.00128.L	&6.2.1-7&	6	&	0.0459	&	0.138	&	...	&	1.4	&	0.2	\\
2MASS J16163345-2521505	&	2021.1.00128.L	&6.2.1-7&	6	&	0.0459	&	0.138	&	2.9	&	1.8	&	0.08	\\
\hline
\end{tabular}
\label{tab:alma_data}
\end{table*}
\FloatBarrier

\FloatBarrier
\begin{table*}[h]
\caption{Output parameters. "Sil" as error in Column 7 means that $a_{\rm min}$(inner) is where temperature of grains is equal to the sublimation temperature of silicates.}
\tiny
\centering
\begin{tabular}{lccccccccc}
\hline
\hline
Star	&	$\chi^2$(sed)	&	$\chi^2$(Alma)	&	\teff	&	$R_*$	& $\log{\dot{M}}$	& $a_{\rm min,inner}$	&	$a_{\rm max,inner}$	&	$M_{\rm inner}$	&	$M_{\rm outer}$		\\
&&& (K) & (\RSun) & (\Msun~$yr^{-1})$ & (au) & (au) & (10$^{-5}$\, \Msun) & (10$^{-3}$\, \Msun)\\
\hline
CW Tau	&	57.1	&	87.2	&	3668	&	1.95	&	-7.53$\pm$0.11	&	0.028$^a$	&	3.2	&	23$\pm$1		&	5.2$\pm$0.3		\\
CX Tau	&	25.9	&	29.1	&	3892$\pm$28	&	1.34$\pm$0.06	&	-9.65$\pm$0.11	&	0.170$\pm$0.010	&	1.6	&	2.2$\pm$0.1		&	1.58$\pm$0.08		\\
CY Tau	&	77.8	&	106.8	&	3752$\pm$29	&	1.27$\pm$0.08	&	-9.10$\pm$0.12	&	0.018$^a$	&	1.4	&	0.71$\pm$0.04	&	10.3$\pm$0.5	\\
FT Tau	&	38.8	&	9.3	&	3313$\pm$24	&	1.56$\pm$0.08	&	-10.20$\pm$0.11	&	0.017$^a$	&	2.5	&	0.88$\pm$0.04	&	16.7$\pm$0.9	\\
DM Tau	&	15.0	&	27.0	&	3572$\pm$29	&	1.20$\pm$0.05	&	-9.08$\pm$0.11	&	0.553$\pm$0.030	&	4.6	&	23$\pm$1		&	32$\pm$2		\\
HO Tau	&	15.7	&	4.5	&	3535$\pm$20	&	1.10$\pm$0.05	&	-9.40$\pm$0.10	&	0.020$\pm$0.001	&	1.07	&	0.33$\pm$0.02	&	11.9$\pm$0.7	\\
DO Tau	&	23.7	&	34.3	&	3533$\pm$25	&	2.10$\pm$0.10	&	-9.07$\pm$0.11	&	0.026$^a$	&	2.0	&	13.9$\pm$0.8		&	14.7$\pm$0.8		\\
Haro 6-33	&	29.9	&	20.7	&	3651$\pm$28	&	1.08$\pm$0.06	&	-9.87$\pm$0.11	&		0.014$^a$	&	1.12	&	45$\pm$2	&	32$\pm$2		\\
HM Lup	&	16.9	&	12.2	&	3501$\pm$19	&	1.36$\pm$0.05	&	-9.41$\pm$0.09	&	0.066$\pm$0.004	&	1.3	&	1.8$\pm$0.11	&	24$\pm$1		\\
Sz 73	&	22.2	&	10.2	&	4011$\pm$29	&	1.11$\pm$0.06	&	-9.56$\pm$0.12	&	0.018$^a$	&	1.4	&	6.3$\pm$0.4	&	23$\pm$1	\\
J16000236-4222145	&	39.0	&	8.9	&	3000$\pm$31	&	1.35$\pm$0.06	&	-11.95$\pm$0.37	&	0.041$\pm$0.006	&	0.94	&	0.26$\pm$0.05	&	11$\pm$3		\\
Sz 91	&	80.0	&	16.1	&	3967$\pm$31	&	1.05$\pm$0.06	&	-9.77$\pm$0.12	&	1.037$\pm$0.058	&	11	&	5.2$\pm$0.3	&	11.7$\pm$0.7	\\
J16085324-3914401	&	72.1&	21.7	&	3302$\pm$14	&	1.42$\pm$0.07	&	-10.60$\pm$0.12	&	0.079$\pm$0.005	&	1.20	&	0.80$\pm$0.04	&	6.9$\pm$0.4	\\
Sz 111	&	11.9	&	12.5	&	3670$\pm$27	&	1.23$\pm$0.04	&	-9.46$\pm$0.11	&	5.9$\pm$0.3	&	21	&	1510$\pm$90	&	21$\pm$1	\\
Sz 114	&	63.8	&	85.5	&	3528$\pm$24	&	1.22$\pm$0.08	&	-9.37$\pm$0.12	&	0.036$\pm$0.002	&	1.23	&	1.46$\pm$0.09	&	9.3$\pm$0.5	\\
Sz 123	&	24.9	&	9.9	&	3542$\pm$25	&	1.17$\pm$0.05	&	-9.50$\pm$0.12	&	0.041$\pm$0.003	&	1.8	&	0.66$\pm$0.04		&	17.9$\pm$0.9		\\
J16145026-2332397	&	35.8	&	22.8	&	3390$\pm$29	&	0.90$\pm$0.05	&	-9.35$\pm$0.11	&	0.010$^a$	&	0.81	&	0.24$\pm$0.01	&	4.7$\pm$0.3	\\
J16163345-2521505	&	47.7	&	19.7	&	3815$\pm$27	&	0.90$\pm$0.04	&	-12.87$\pm$0.11	&	0.076$\pm$0.004	&	1.01	&	0.84$\pm$0.05	&	1.95$\pm$0.11	\\
\hline
\end{tabular}
\tablefoot{$^a$ $a_{\rm min}$(inner) corresponds to the value where temperature of grains is equal to the sublimation temperature of silicates. }
\label{tab:output_data}
\end{table*}
\FloatBarrier

\FloatBarrier
\begin{table*}[h!]
\caption{Derived quantities.}
\centering
\tiny
\begin{tabular}{lcccccccl}
\hline
\hline
Star	&$R_{\rm max,grain,inner}$	 & Slope $\beta$ &$M_{\rm dust}$(inner)	&	$M_{\rm dust}$(outer)	&	Age	&	Age lit & ref&Comments	\\
& (cm) && (M$_{\rm Earth})$& (M$_{\rm Earth})$& (Myr)& (Myr)& \\
\hline
CW Tau	&	0.73	$\pm$	0.14	& -2.845$\pm$0.008&	0.032	&	17	&	2.0 &	1.8	&	3	& High veiling \\
CX Tau	&	1.4	$\pm$	0.2	& -2.795$\pm$0.008&	0.0099	&	5.2	&	2.2	$\pm$	0.4	&	1.8	&	3	&		\\
CY Tau	&	0.10	$\pm$	0.02	& -2.843$\pm$0.009&	0.00016	&	34	&	2.0	$\pm$	0.3	&	1.8	&	3	&		\\
FT Tau	&	3.0$\pm$	0.6	& -2.810$\pm$0.018&	0.0090 &	55	&	1.4	$\pm$	0.2	&	0.7	&	2	&		\\
DM Tau	&	0.041	$\pm$	0.006	& -2.800$\pm$0.008&	0.00009	&	104	&	3.5	$\pm$	0.6	&	2.8	&	6 	&		\\
HO Tau	&	1.9	$\pm$	0.3	& -2.811$\pm$0.008&	0.0020	&	39	&	4.3	$\pm$	0.6	&	1.7	&	7 	&	Central peak 	\\
DO Tau	&	1.1	$\pm$	0.2	& -2.810$\pm$0.018 &	0.048	&	48	&	0.8	$\pm$	0.1	&	2.3	&	1		&	Central peak\\
Haro 6-33	&	2.5	$\pm$	0.5	& -2.829$\pm$0.008&	0.47	&	104	&	3.6	$\pm$	0.6	&	1.9	&	11	&		\\
HM Lup	&	0.9	$\pm$	0.1	& -2.831$\pm$0.008&	0.0048	&	79	&	2.1	$\pm$	0.2	&	2.9	&	5	&		\\
Sz 73	&	0.9	$\pm$	0.2	& -2.838$\pm$0.009&	0.016	&	77	&	10	$\pm$	2	&	9.0	&	10	&	Central peak	\\
J16000236-4222145	&	0.4	$\pm$	0.2	& -2.811$\pm$0.007&	0.00029	&	37	&	2.7	$\pm$	0.4	&	1.5	&	5	&	Central peak	\\
Sz 91	&	0.0048	$\pm$	0.0006	& -2.888$\pm$0.009&	0.0	&	39	&	6	$\pm$	2	&	4.0	&	8	&		\\
J16085324-3914401	&	0.29	$\pm$	0.05	& -2.830$\pm$0.007&	0.0005	&	23	&	2.0	$\pm$	0.4	&	3.1	&	4 	&	Central peak	\\
Sz 111	&	0.017	$\pm$	0.003	& -2.940$\pm$0.008&	0.0023	&	690	&	4.6	$\pm$	0.7	&	4.2	&	5	&		\\
Sz 114	&	1.1	$\pm$	0.2	& -2.830$\pm$0.008&	0.0050	&	31	&	2.2	$\pm$	0.4	&	2.4	&	5 	&	Central peak	\\
Sz 123	&	1.7	$\pm$	0.3	& -2.813$\pm$0.008&	0.0024	&	59	&	4.0	$\pm$	1.2	&	4.4	&	5	&		\\
J16145026-2332397	&	0.44	$\pm$	0.08	& -2.809$\pm$0.008&	0.00027	&	15.5	&	7.5	$\pm$	1.3	&	7.1	&	9 	&	Central peak	\\
J16163345-2521505	&	2.5	$\pm$	0.4	& -2.805$\pm$0.008&	0.0066	&	6.4	&	10	$\pm$	2	&	7.5	&	9	&		\\
\hline
\end{tabular}
\tablebib{
(1) Age of L1527 (Gratton et al, in prep.);
(2) Age of L1558 (Gratton et al, in prep.);
(3) Age of L1529 (Gratton et al, in prep.);
(4) \citet{Galli2020};
(5) \citet{Frasca2017};
(6) Age of L1529 (Gratton et al, in prep.);
(7) Age of L1529 (Gratton et al, in prep.);
(8) Maio et al. 2026, in prep.;
(9) \citet{Ratzenbock2023};
(10) \citet{Alcala2014};
(11) \citet{Connelley2026}.}
\label{tab:derived_data}
\end{table*}
\FloatBarrier

\section{Fits to ALMA data and SED}
\label{sec:fit}

\begin{figure}[h]
    \includegraphics[width=0.5\textwidth]{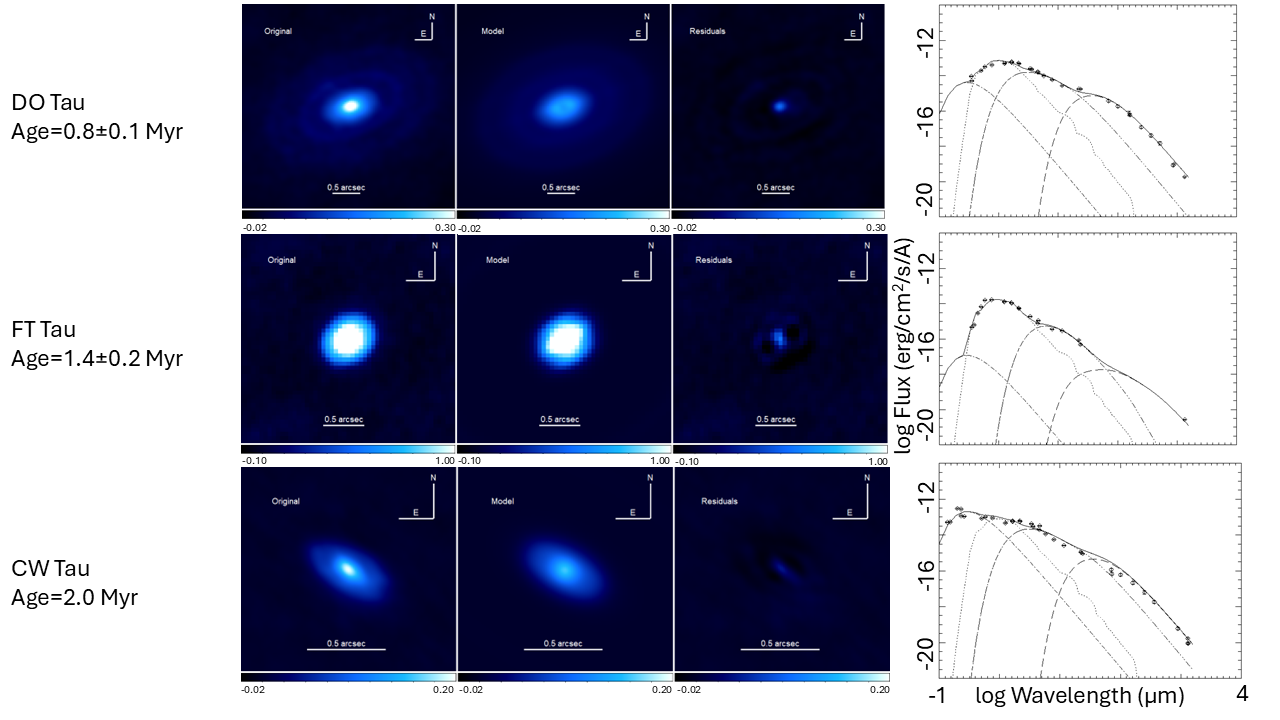}
    \includegraphics[width=0.5\textwidth]{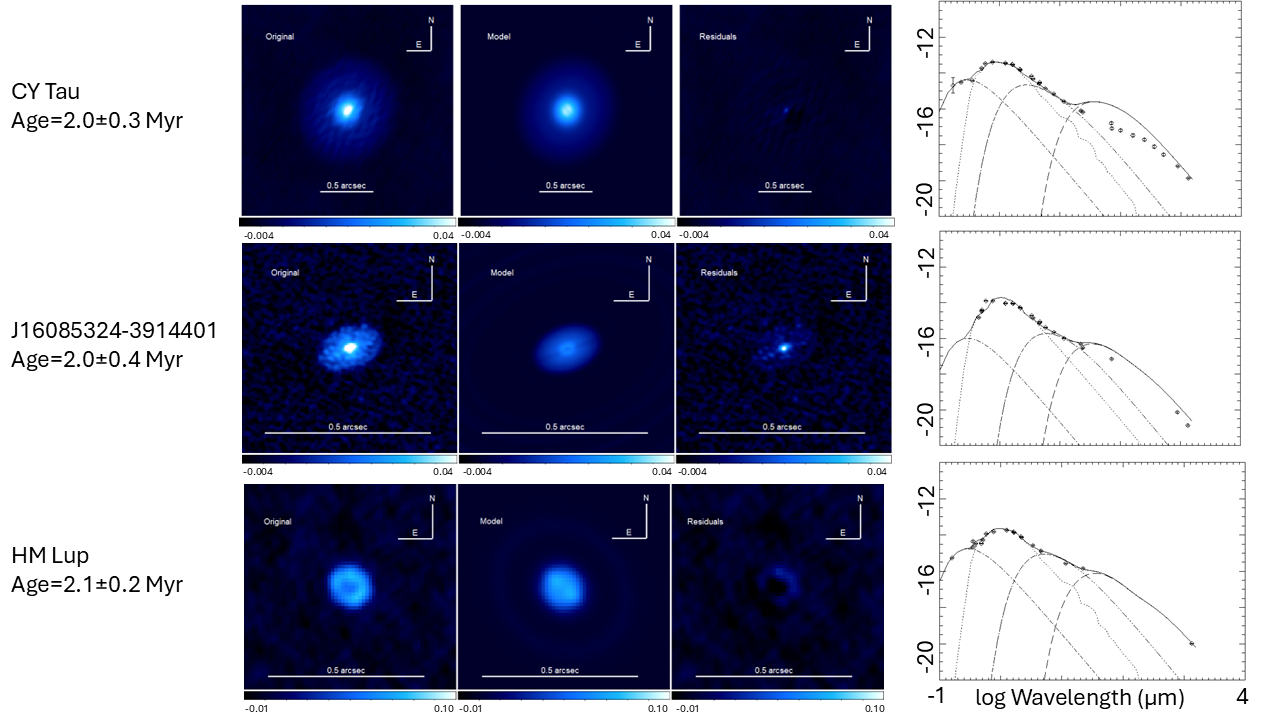}
    \caption{From top to bottom: results obtained with the model presented in this work for DO Tau, FT Tau, CW Tau, CY Tau, 2MASS J16085324-3914401, and HM Lup. From left to right: the original ALMA continuum image; the model of the ALMA image; the residuals after subtraction of the model from the original data, and the SED. In the last panel symbols with error bars are observations and the solid line is the fit. Individual contributions by the accretion region, the stellar photosphere, the inner disc, and the outer disc are shown as dot-dashed, dotted, long dashed and short dashed lines, respectively. }
    \label{fig:alma_sed_a}
\end{figure}

\begin{figure}[h]
    \centering
    \includegraphics[width=0.5\textwidth]{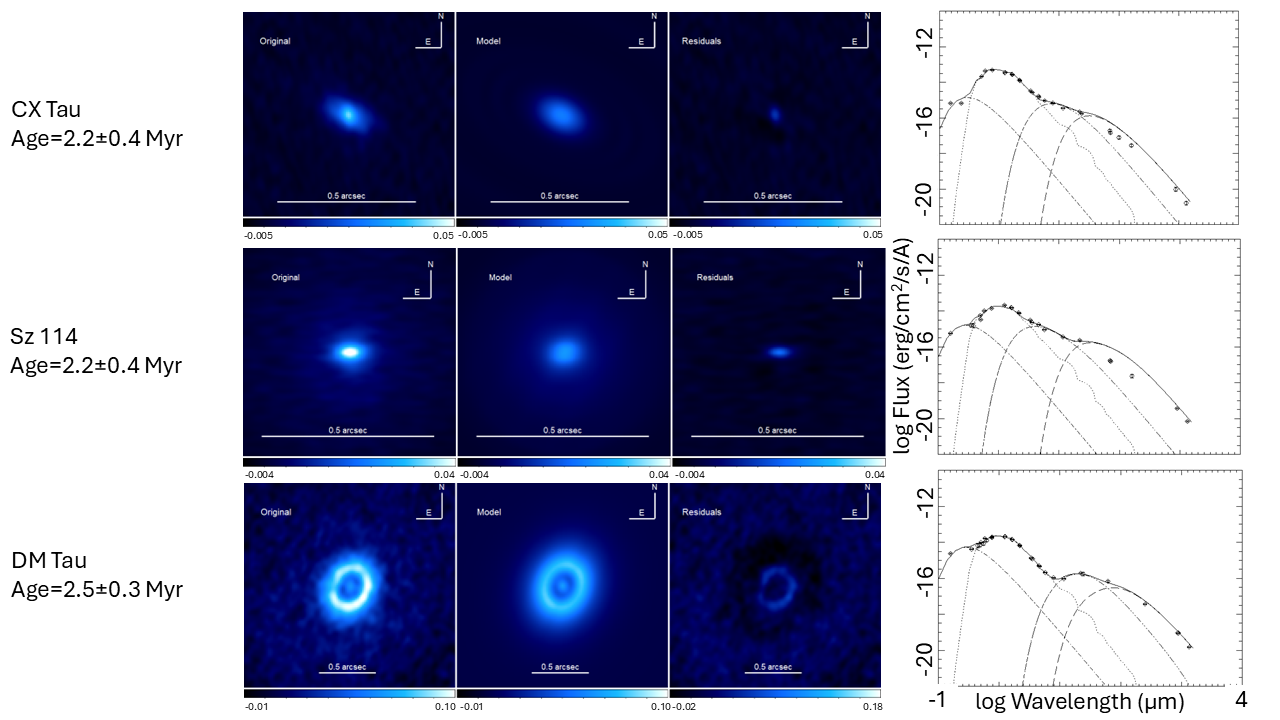}
    \includegraphics[width=0.5\textwidth]{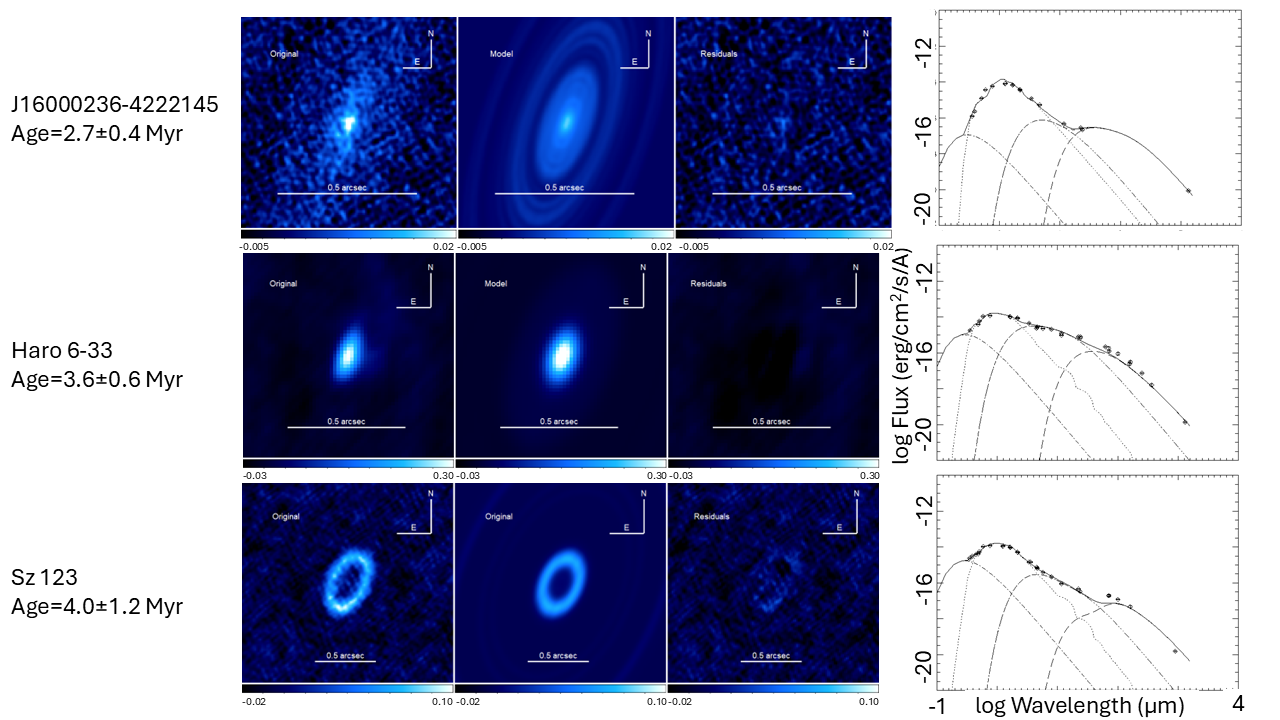}
    \caption{Same as Fig.~\ref{fig:alma_sed_a}, but for CX Tau, Sz 114, DM Tau, 2MASS J16000236-4222145, Haro 6-33, and Sz 123 }
    \label{fig:alma_sed_b}
\end{figure}

\begin{figure}[t]
    \centering
    \includegraphics[width=0.5\textwidth]{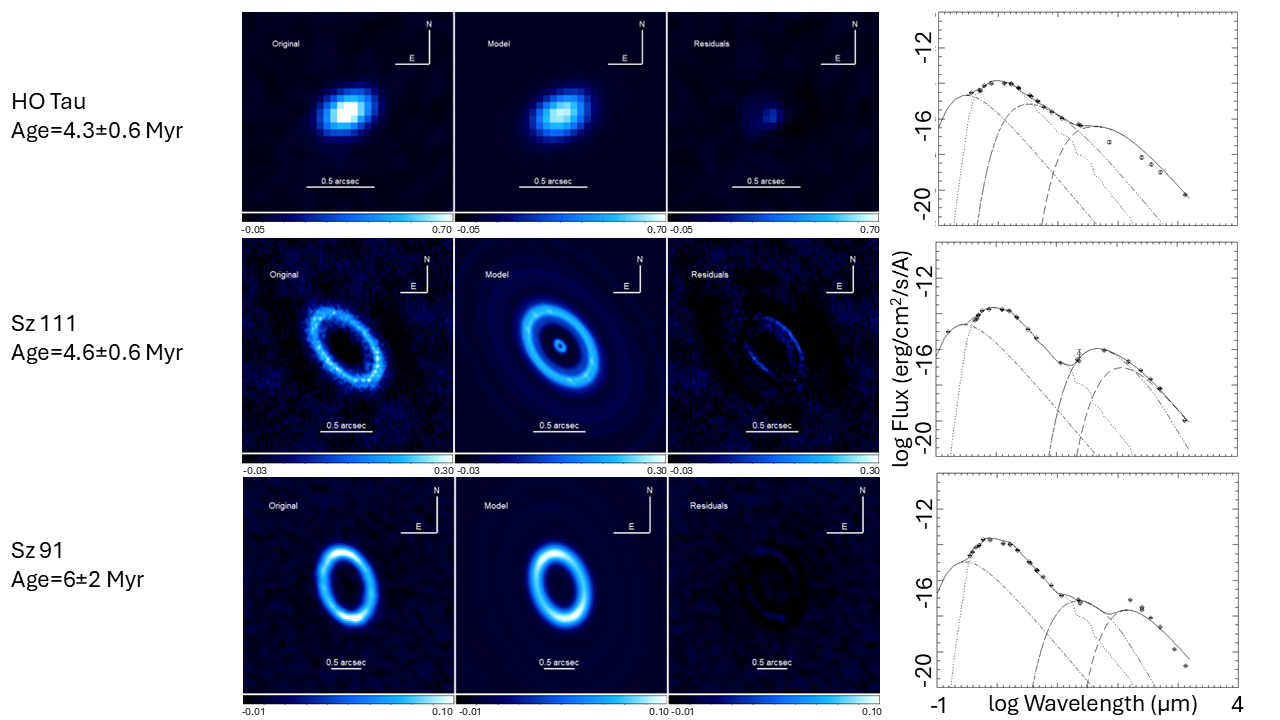}
    \includegraphics[width=0.5\textwidth]{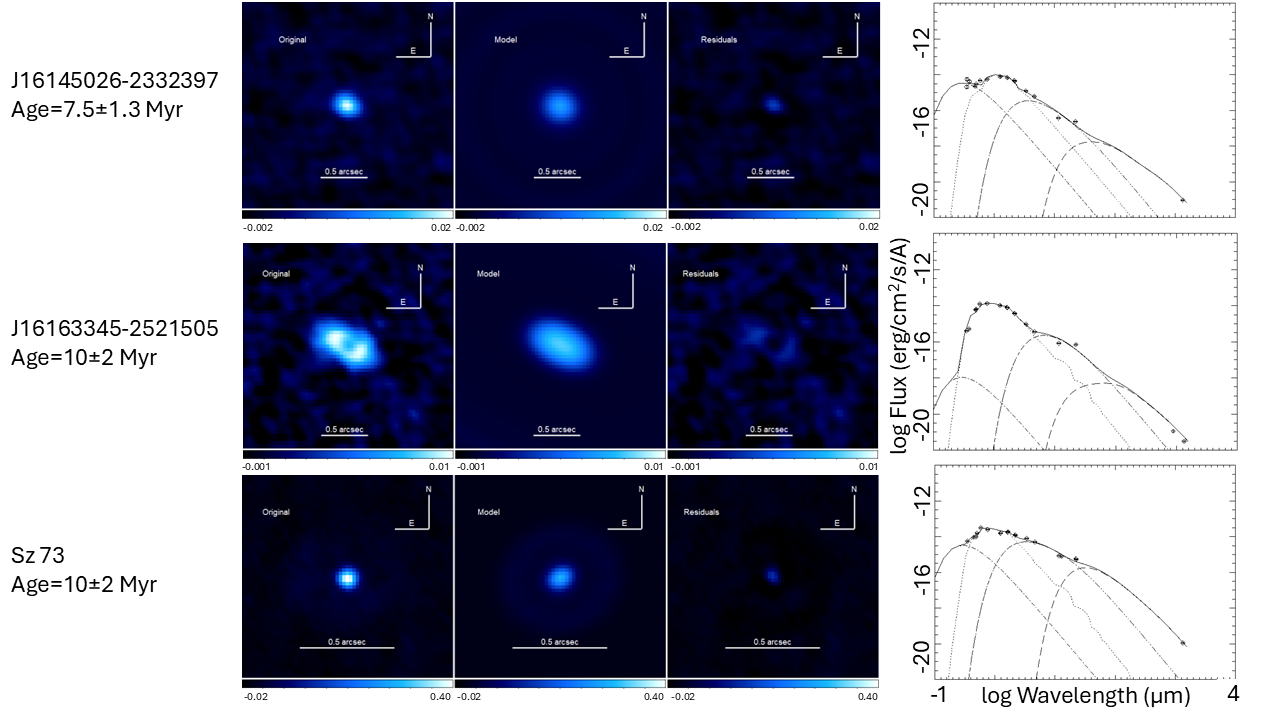}
    \caption{Same as Fig.~\ref{fig:alma_sed_a}, but for HO Tau, Sz 111, Sz 91, 2MASS J16145026-2332397, 2MASS J16163345-2521505, and Sz 73. }
    \label{fig:alma_sed_c}
\end{figure}

\section{Disc surface density profiles}
\label{sec:profile}

\begin{figure}[h]
    \centering
    \includegraphics[width=0.4\textwidth]{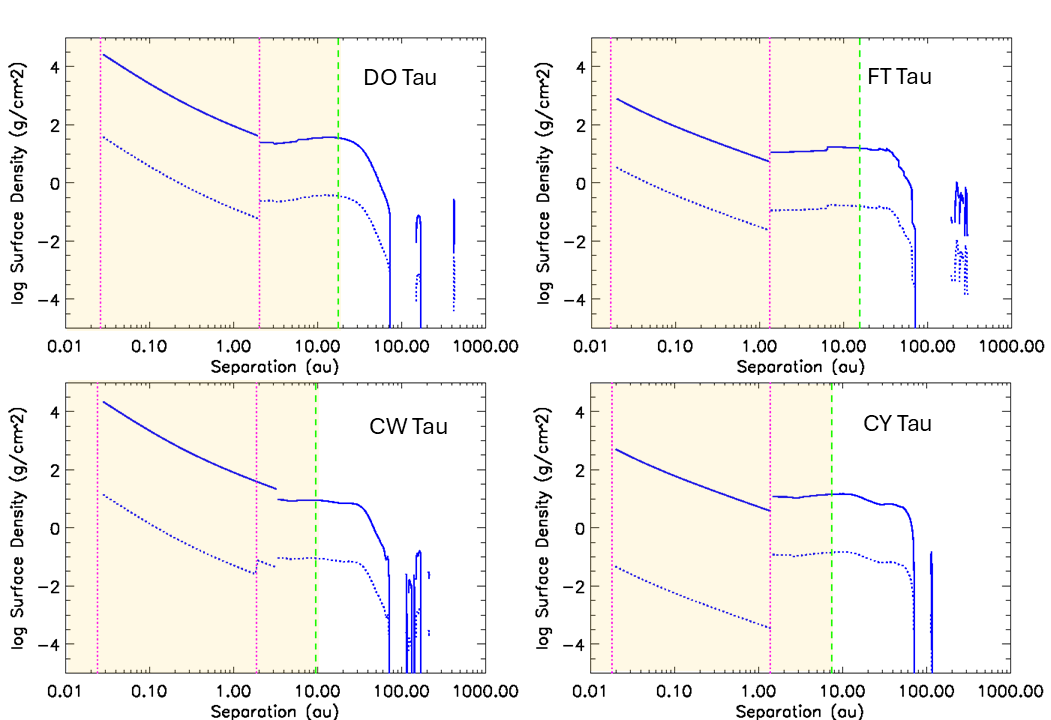}
    \includegraphics[width=0.4\textwidth]{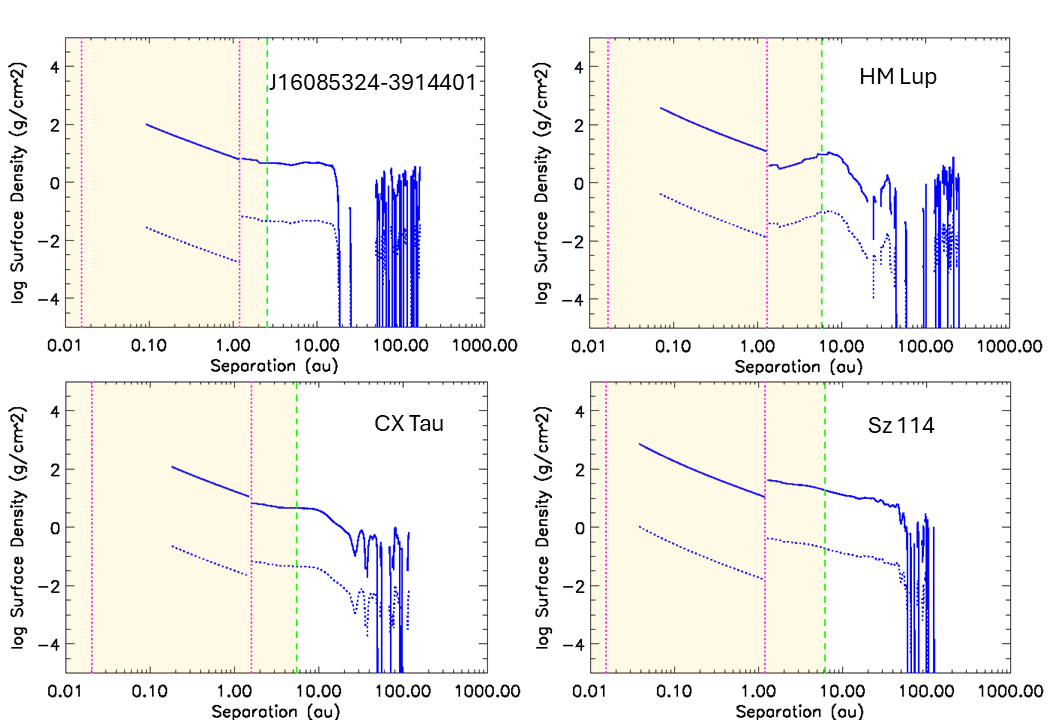}
    \caption{Radial surface density profiles from our model. Solid lines are the gas surface density; dashed lines are the surface density in grains. Vertical magenta dotted lines mark the location of the sublimation distances for silicates and ice. The vertical green dashed line represents the resolution of the ALMA images and the shaded area is the region unresolved in these images.}
    \label{fig:profile_a}
\end{figure}

\begin{figure}[h]
    \centering
    \includegraphics[width=0.4\textwidth]{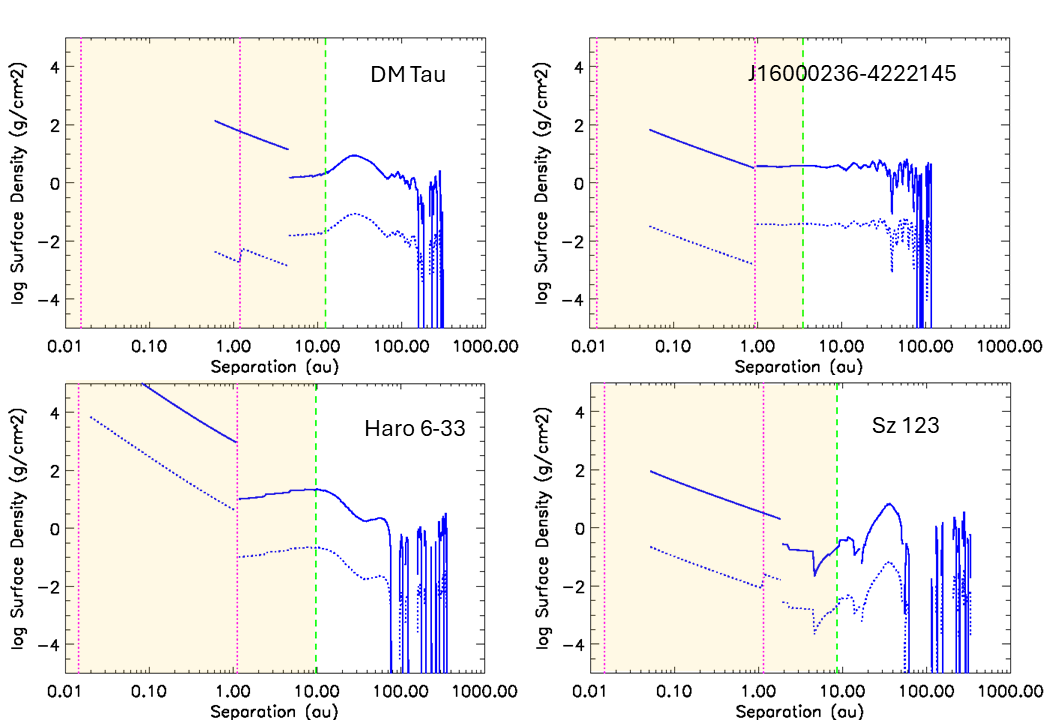}
    \includegraphics[width=0.4\textwidth]{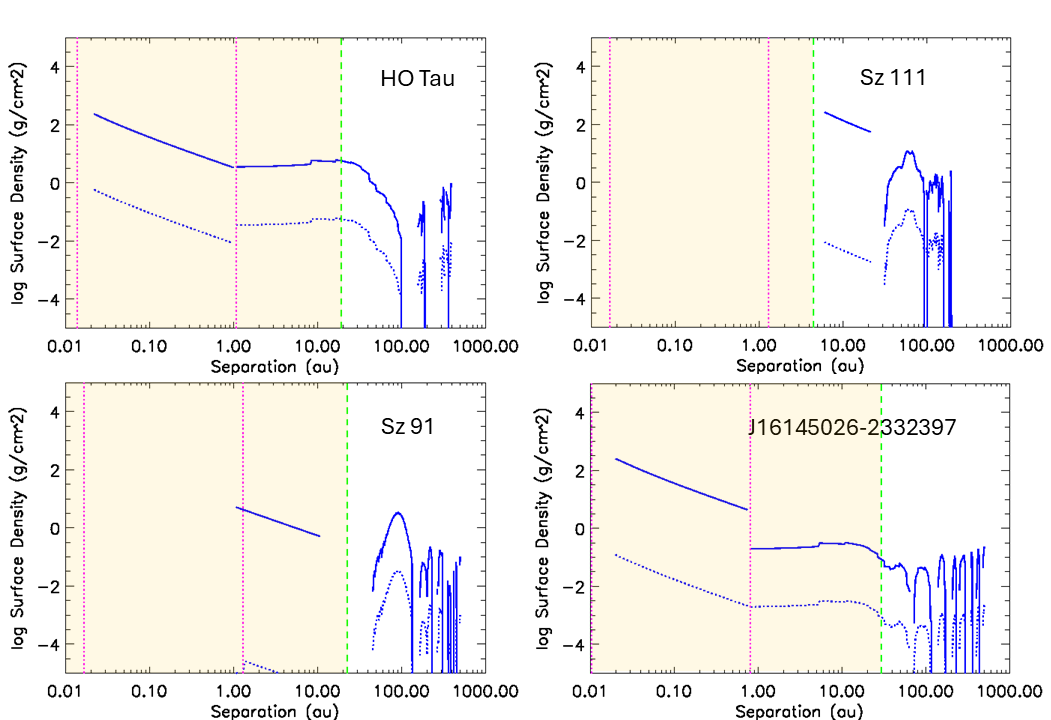}
    \includegraphics[width=0.4\textwidth]{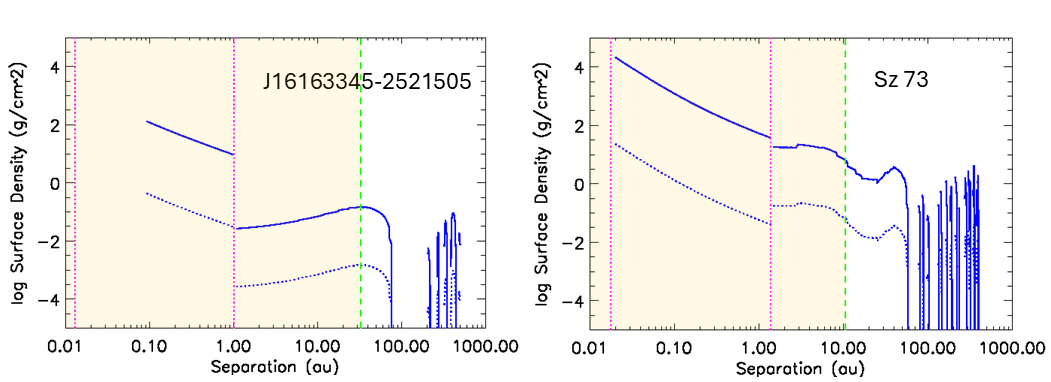}
    \caption{Same as Fig.~\ref{fig:profile_a} but for DMTau, J16000236-4222145, Haro 6-33, Sz123, HOTau, Sz111, Sz91, J16145026-2332397, J16163345-2521505 and Sz73.}
    \label{fig:profile_b}
\end{figure}

\newpage

\section{Corner plots}
\label{sec:corner_plot}

\begin{figure}[h]
    \centering
    \includegraphics[width=\columnwidth]{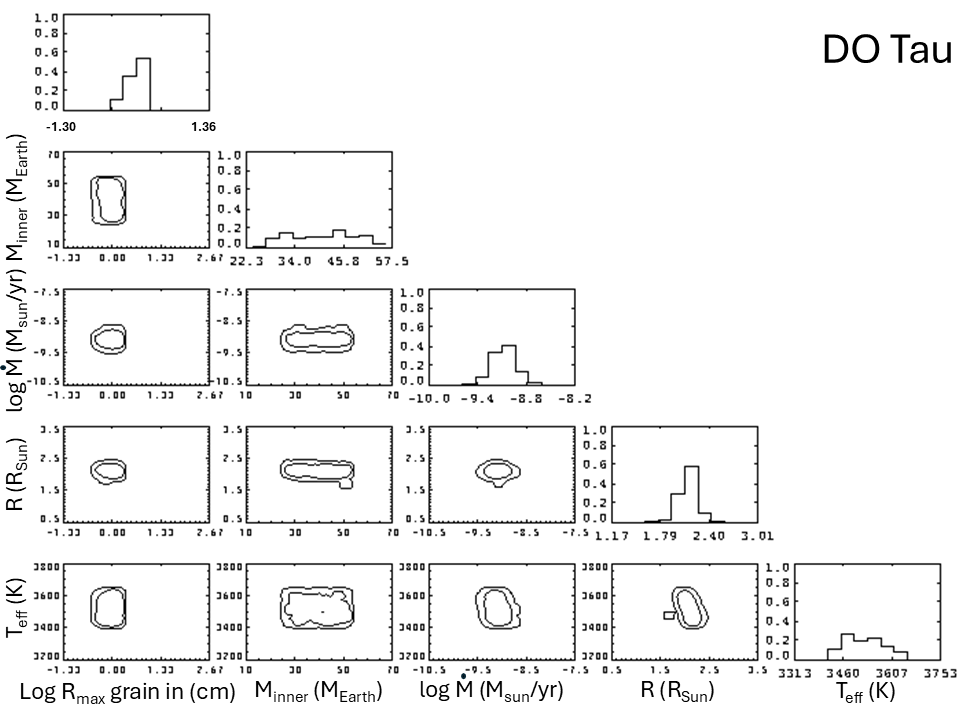}
    \includegraphics[width=\columnwidth]{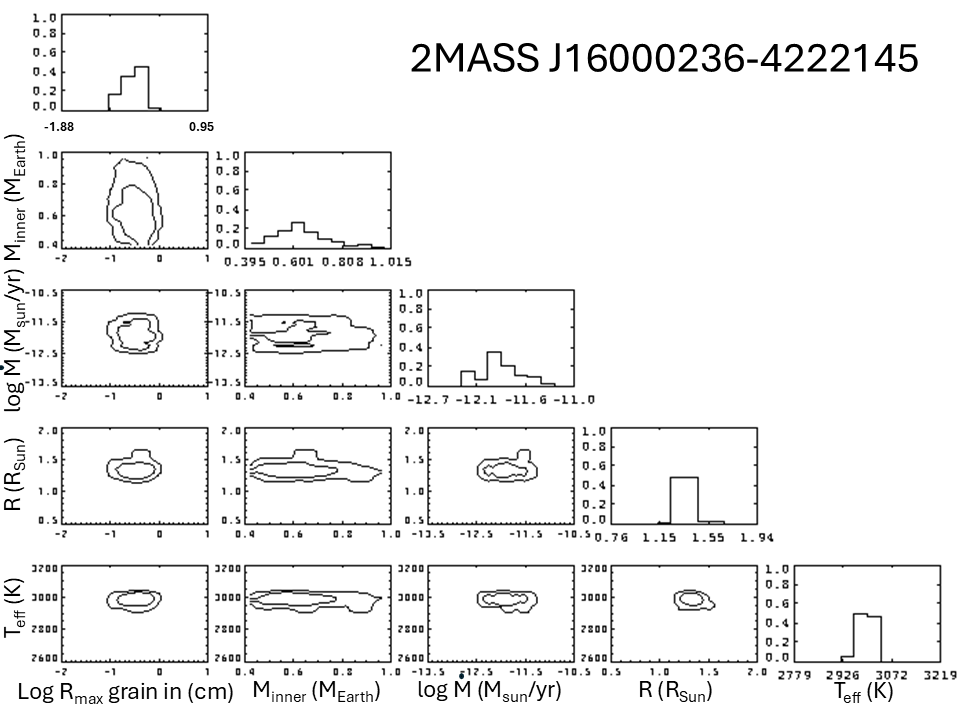}
    \includegraphics[width=\columnwidth]{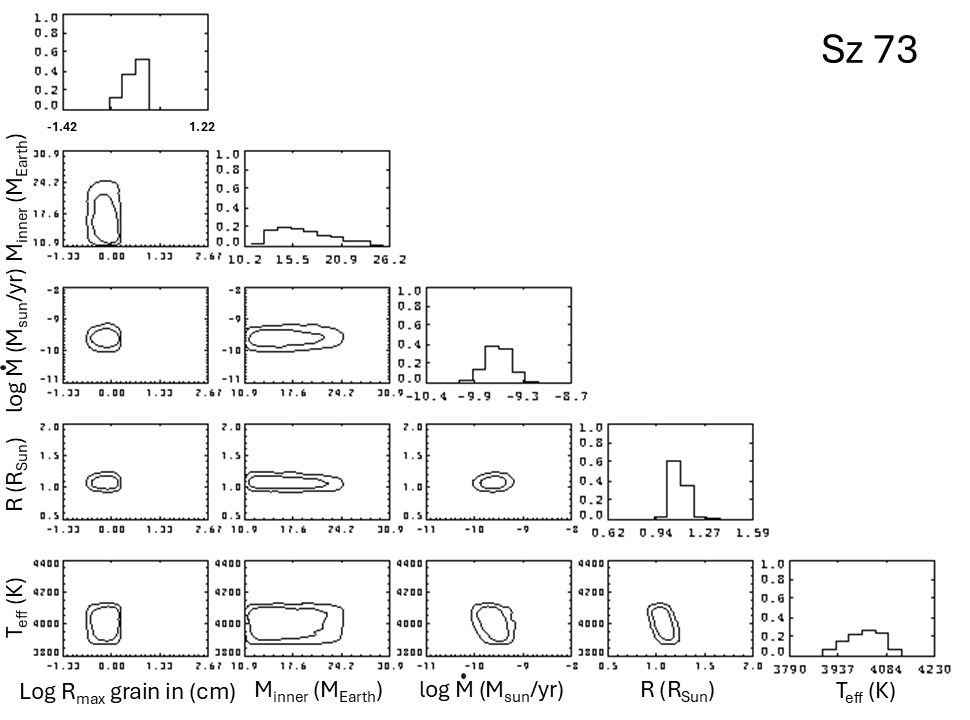}
    \caption{Corner plots showing the correlations between parameters for DO Tau (upper panel), 2MASS J16000236-4222145 (middle panel), and Sz 73  (lower panel). that have probabilities 68.3\% (1$\sigma$) and 95.5\% (2$\sigma$) of being acceptable. }
    \label{fig:corner_plot}
\end{figure}

\end{appendix}
\end{document}